\documentclass[final,10pt,journal,twocolumn]{IEEEtran}

\input{def.tex}

\begin{document}

\title{Impact of General Channel Aging Conditions on the Downlink Performance of Massive MIMO}
\author{
Anastasios K. Papazafeiropoulos, \emph{Member, IEEE}
\thanks{Copyright (c) 2015 IEEE. Personal use of this material is permitted. However, permission to use this material for any other purposes must be obtained from the IEEE by sending a request to pubs-permissions@ieee.org.}
\thanks{
        Anastasios K. Papazafeiropoulos is with the Department of Electrical and Electronic Engineering, Imperial College London, SW7 2AZ  London, United Kingdom (e-mail:a.papazafeiropoulos@imperial.ac.uk). }
\thanks{Parts of this work are accepted in  IEEE Global Communications Conference (GLOBECOM), San Diego, USA, Dec. 2015~\cite{Papazafeiropoulos2015b}.}}
\maketitle 
\begin{abstract}
Recent works have identified massive multiple-input-multiple-output (MIMO) as a key technology for achieving substantial gains in spectral and energy efficiency. Additionally, the turn to low-cost transceivers, being prone to hardware impairments is the most effective and attractive way for cost-efficient applications concerning massive MIMO systems.  In this context, the impact of channel aging, which severely affects the performance, is investigated herein by considering a generalized model. Specifically, we show that both Doppler shift because of the users' relative movement as well as phase noise due to noisy local oscillators (LOs)  contribute to channel aging. To this end, we first propose a joint model, encompassing both effects, in order to investigate the performance of a massive MIMO system based on the inevitable time-varying nature of realistic mobile communications. Then, we derive the deterministic equivalents (DEs) for the signal-to-noise-and-interference ratios (SINRs) with maximum ratio transmission (MRT) and regularized zero-forcing precoding (RZF). Our analysis not only demonstrates a performance comparison between MRT and RZF under these conditions, but most importantly, it reveals interesting properties regarding the effects of user mobility and phase noise. In particular, the large antenna limit behavior depends profoundly on both effects, but the burden due to user mobility is much more detrimental than phase noise even for moderate user velocities ($\approx30$ km/h), while the negative impact of phase noise is noteworthy at lower mobility conditions. Moreover, massive MIMO systems are favorable even in general channel aging conditions. Nevertheless, we demonstrate that  the transmit power of each user to maintain a certain quality of service can be scaled down at most by $1\sqrt{M}$ ($M$ is the number of BS antennas), which indicates that the joint effects of phase noise and user mobility do not degrade the power scaling law, but only the achievable sum-rate.
\end{abstract}
\section{Introduction} 
The increasing demand for wireless services and higher data volume per area has brought to the research forefront of both the scientific community (e.g., the METIS project~\cite{METIS}) and the standardization bodies  a new technology known as massive multiple-input-multiple-output (MIMO) networks~\cite{Marzetta2010,Rusek2013}. The keystone of this concept, currently considered for the design of  next generation wireless networks~\cite{5g,Larsson2014,Osseiran2014}, is based on that the numbers of base station (BS) antennas and users are scaled up by 1-2 orders of magnitude under the same time-frequency resources.  This scientific area is motivated by the resulting gains stemming from large channel matrices, i.e., the arising advantages due to the asymptotics of random matrix theory~\cite{Couillet2011}. Specifically, by  allowing the number of BS antennas tend to infinity,  small-scale fading and thermal noise average out, as well as the analysis is simplified. 


Notably, the channel state information (CSI)  is  pivotal to the performance of  such systems that take into account for multi-user MIMO (MU-MIMO)~\cite{Gesbert2007}. Thus, recognizing the need for a realistic study with imperfect CSI, we focus on an important factor-phenomenon, being present in time-varying channels, that deteriorates the quality of CSI available at the BS in 
time-division-duplex (TDD) systems. Our consideration includes TDD systems and not frequency-division-duplex (FDD) systems, since the  application of the latter in massive MIMO  systems is meaningful only when  
the channel matrices are sparse~\cite{Bjoernson2015}\footnote{The hypothesis of sparsity is still an open question requiring further research and channel measurements.}. This phenomenon, known as {\em  channel aging}, describes the mismatch appearing between the current  and the estimated channel due to the relative movement between the users and the BS antennas in addition to any processing delays. In other words, it models the divergence occurring between the channel estimate used in the precoder/detector, and the channel over which the data transmission actually  takes place. 

Although the significance of  channel aging due to user mobility, its effect on the performance of massive MIMO systems has not been addressed adequately as  few studies  on this area exist. In fact, only the user mobility has been considered as the only channel aging cause~\cite{Truong2013,Papazafeiropoulos2014,Papazafeiropoulos2014WCNC,Papazafeiropoulos2015a,Papazafeiropoulos2015,PapazafeiropoulosPIMRC, Kong2015,Kong2015a}. Basically, channel aging  due to only user mobility was the focal point in~\cite{Truong2013}, where an application of the deterministic equivalent (DE) analysis was presented by considering linear techniques in the uplink and downlink in terms of maximum ratio combining (MRC) detector and  maximum ratio transmission (MRT) precoder, respectively. Further investigations have been performed in~\cite{Papazafeiropoulos2015a}, where more sophisticated linear techniques are employed, i.e., minimum-mean-square-error (MMSE) receivers (uplink) and regularized zero-forcing precoders (downlink) are considered, and a comparison between the various linear transceiver techniques is performed. Moreover, in~\cite{Papazafeiropoulos2015}, the optimal linear receiver in the case of cellular massive MIMO  systems has been derived  by exploiting the correlation between the channel estimates and the interference from other cells, while in~\cite{PapazafeiropoulosPIMRC}, the uplink analysis of a cellular network with zero-forcing (ZF) receivers that holds for any finite as well as infinite number of BS antennas has been provided.

In general, hardware undergoes several types of impairments such as high power amplifier nonlinearities and I/Q imbalance~\cite{Schenk2008,Studer2010,Qi2012,Mehrpouyan2012,Goransson2008,Bjornson2012Optimal,Qi2010,Bjoernson2013,Zhang2015,Pitarokoilis2015,Bjornson2015,Krishnan2015,Bjornson2014,Demir2000,Bittner2007,Wu2004,Petrovic2007,Krishnan2014,Pitarokoilis2013}
\footnote{The arising hardware mismatch between the uplink and the downlink has a negative effect on the  channels reciprocity. For example, a recent insightful work (see \cite{Zhang2015} and  references therein) has presented the performance analysis in the large number of antennas regime as well as pre-precoding and post-precoding calibration schemes. Herein, focusing on the fundamentals of channel aging, where we want to quantify the individual contributory time-dependent factors (Doppler shift and phase noise), we neglect any hardware mismatch. However, future work can
investigate this mismatch in the channel aging.}.  These hardware impairments can be classified into two categories. In the first category, the variable, describing the hardware impairments, is multiplied with the channel vector, and it might cause channel attenuations and phase shifts. In the case of slow variation of these impairments, they can be characterized as sufficiently static, and thus, can be assimilated by the channel vector by an appropriate scaling of its covariance matrix or due to the property of circular symmetry of the channel distribution.      Note that this factor cannot be incorporated by the channel vector by an appropriate scaling of its covariance matrix or due to the property of circular symmetry of the channel distribution, when it changes faster than the channel. Focusing on this scenario, we include in our analysis the phase noise in the case which results to its accumulation within the channel coherence period~~\cite{Mehrpouyan2012,Bjornson2015,Pitarokoilis2015,Krishnan2015}. Given that additive transceiver impairments are not time-dependent, we have not included them in our analysis~\cite{Bjoernson2013,Bjornson2014,Bjornson2015}.

Hence, deeper consideration reveals that  user mobility is not the only source of channel aging. Interestingly, {the unavoidable \em phase noise}, being one of the most severe transceiver impairments~\cite{Demir2000}, contributes to the further degradation of the system performance due to channel aging because it accumulates over time. Ideally, local oscillators (LOs), used during the conversion of the baseband signal to bandpass and vice-versa, should output a sinusoid with stable amplitude, frequency, and phase. Unfortunately, the presence of phase noise  is inevitable due to the inherent imperfections in the circuitry of  LOs that bring up to the sinusoid a random phase drift.  A lot of research has been conducted regarding phase noise in conventional MIMO systems.  In fact, a big part of the literature has dealt with multi-carrier MIMO systems that employ orthogonal frequency-division multiplexing (OFDM)~\cite{Bittner2007,Wu2004,Petrovic2007,Krishnan2014}. For example, in~\cite{Wu2004}, the signal-to-interference-and-noise-ratio (SINR) degradation in OFDM is studied, and a method to mitigate the effect of phase noise is proposed. In~\cite{Mehrpouyan2012}, a method to jointly estimate the channel coefficients and the phase noise in a point-to-point MIMO system is presented. Not to mention that the degradation appears to be more significant in coherent communications such as massive MIMO  systems.  Specifically, although massive MIMO technology is attractive (cost-efficient) for network deployment, only if the antenna elements consist of inexpensive hardware (more imperfections), most of the research contributions are based on the strong assumption of perfect hardware without phase noise, which is quite idealistic in practice. Reasonably, the introduction of a larger number of antennas will have a more severe contribution to the degradation provoked by phase noise, which could be partially avoided in conventional MIMO  systems, where more expensive hardware with less induced phase is affordable. Hence, the  larger the number of antennas and the corresponding employed LOs, the larger the impact of phase noise. Thus, in the case of massive MIMO  systems, it is a dire necessity to study the impact of channel aging induced by phase noise. Furthermore, motivated by the need of employing low-quality RF elements (e.g., oscillators) for cost-effective construction of large BS antenna arrays, it is  expected that the phase noise will be more severe in a massive MIMO setting.  In particular, the large number of antennas regime has been partially     investigated~\cite{Pitarokoilis2013,Pitarokoilis2015,Bjornson2015,Krishnan2015}. Specifically, the phase noise has been taken into account for single carrier  uplink  massive MIMO by considering ZF and time-reversal MRC  in~\cite{Pitarokoilis2013} and~\cite{Pitarokoilis2015}, respectively. 
Regarding the downlink,  an effort to address the effect of phase noise, being of great importance in massive MIMO systems,  has taken place only in~\cite{Krishnan2015}. However, the analysis therein does not account for fast time-varying conditions due to the mobility of the users.

In this paper, we provide a novel realistic CSI model that allows studying channel aging  thoroughly by means of the simultaneous impact of    user mobility and phase noise at the LOs. Taking into account that   next generation wireless systems need to be cost-efficient, implying that real systems may include low-quality RF elements with more imperfections, makes the significance of the proposed model in describing the ineluctable time-varying nature of realistic scenarios remarkable.  Under these conditions, we analyze the downlink performance of a massive MIMO system with imperfect CSI using linear precoders (MRT and RZF). To this end, we invoke random matrix theory (RMT) tools, and we derive the DEs (asymptotically tight approximations as system's dimensions increase). Actually, these approximations agree with simulation results as has been  already shown in the literature~\cite{Couillet2011,Truong2013,Papazafeiropoulos2015a}. As a result, these deterministic approximations make any lengthy simulations
unnecessary. Furthermore, as far as phase noise is concerned, we address two system arrangements: a) all BS antennas are connected to a common LO (CLO), and b) each BS antenna includes its own separate LO (SLOs setup). The contributions of this paper can be summarized as follows:
\begin{itemize}
 \item Given a single cell MU-MIMO realistic system model with path loss and shadowing, we present a joint channel-phase noise estimate describing the impact of channel aging by incorporating the time-varying effects of Doppler shift and phase noise at the BS and user elements (UEs).
 \item Using tools of large RMT, we obtain the DEs of the downlink SINRs, when the BS applies MRT and RZF precoders under imperfect CSI due to uplink training and time variation of the channel (generalized channel aging). Moreover, we account for special cases of the channel covariance, corresponding to various interesting effects  described by the large-scale component, that simplify a lot the resulting expressions.
 \item We demonstrate that the square root power-law, generally holding in imperfect CSI scenarios, stands for the case of generalized channel aging as well. Hence, the transmit power per user can be cut down by scaling the number of BS antennas to the inverse of their square root.
 \item We corroborate our analysis concerning deterministic approximations by performing simulations coinciding with the analytical expressions.  
 \item We elaborate on insightful conclusions, drawn from our analysis. For example, Doppler shift has a far more severe impact than phase noise even for low vehicle velocities. 
In addition, massive MIMO  systems outperform conventional MIMO even in generalized channel aging conditions with RZF precoder behaving better than MRC for various Doppler shifts and phase noise severities.  
 \end{itemize}

The remainder of this paper is organized as follows. Section II introduces the system  and signal models accounting for both the effects of user mobility and phase noise, which allow a more complete characterization of channel aging. Next, Section III presents the downlink transmission, in terms of the received signal, the applied precoders, and the achievable user rate.  In Section IV, we proceed with the main results that include the derivation of the DEs of the downlink SINRs, as well as the investigation of certain simpler cases such as the scenario where the large-scale fading component is absent. Section V reports a set of numerical results that confirms the validity of our analytical results, and sheds light on the behavior of channel aging for various mobility environments, phase noise settings, and a number of antennas in a massive MIMO system. We summarize our results and observations in Section VI. In Appendix A, some useful results from the literature are presented, while the following appendices provide the proofs for the main analytical results.

\textit{Notation:} Vectors and matrices are represented by boldface lower and upper case symbols. $(\cdot)^\T$, $(\cdot)^\H$, and $\tr{\cdot}$ denote the transpose, Hermitian  transpose, and trace operators, respectively. The expectation and variance operators, as well as the spectral norm of a matrix are denoted by $\EE\left[\cdot\right]$ and $\mathrm{var}\left[\cdot\right]$, as well as $\| \cdot \|$,
respectively, while $\mathrm{Re}\{\cdot\}$  is the real part of a complex number. The  $\mathrm{diag}\{\cdot\}$ operator generates a diagonal matrix from a given vector, and the symbol $\triangleq$ declares definition. The notations $\mathcal{C}^{M}$ and $\mathcal{C}^{M\times N}$ refer to complex $M$-dimensional vectors and  $M\times N$ matrices, respectively. Finally, $\bb \sim \cC\cN{(\b0,\mathbf{\Sigma})}$  or $\bb \sim \cN{(\b0,\mathbf{\Sigma})}$ denote a circularly symmetric complex Gaussian  or a real Gaussian vector $\bb$ with zero-mean and covariance matrix $\mathbf{\Sigma}$.

\section{System Model}
This work considers a single-cell system, where a BS communicates with a number of single-antenna non-cooperative UEs belonging to the set $\mathcal{K}$ with cardinality $K=|\mathcal{K}|$ being the number of UEs. In particular,  each UE is assigned an index $k$ in $\mathcal{K}$, while the BS is deployed with an array of $M$ antennas\footnote{The antennas can be co-located at the BS or can be distributed as fully-coordinated, multiple, smaller BSs~\cite{Bjornson2015}.}. Since each UE has a single antenna, the terms user and user element, are used interchangeably in the following analysis. Our interest focuses on large-scale topologies, where both number of antennas $M$ and users $K$ grow infinitely large while keeping a finite ratio $\beta$, i.e., $M, K\to \infty$ with $K/M=\beta$ fixed such that $\mathrm{lim\,sup}_{M} K/M < \infty$\footnote{ In the following, in order to simplify notation, $M\to \infty$ will be used instead of $M, K\to \infty$, unless  stated otherwise.}.

During the transmission  of the $n$th symbol, the channel gain vector between the BS and user $k$, denoted by $\bh_{k,n}  \in \bbC^{M}$ exhibits flat-fading, and it is assumed constant for the symbol period, while it may vary slowly from symbol to symbol. Basically, given the time-frequency resources of the system, the symbol duration is assumed smaller or equal to the coherence time of all the users. Moreover, in order to account for certain inevitable effects such as path loss and lognormal shadowing, which are dependent on the distance of the users from the BS,  we express the channel between user $k$ and the BS as
\begin{align}
 \bh_{k} = \bR^{1/2}_{k}\bw_{k},
\end{align}
where $\bR_{k} =\mathbb{E}\left[ \bh_{k,n} \bh_{k,n}^{\H}\right]\in \bbC^{M \times M}$ is a deterministic Hermitian-symmetric positive-definite matrix representing the aforementioned effects\footnote{The independence of $\bR_{k}$ from the time index, $n$, originates from the assumption that the  shadowing vary with time in a longer slower pace  than the coherence time~\cite{Truong2013,Papazafeiropoulos2014,Papazafeiropoulos2014WCNC}. Interestingly, this model is quite versatile, since  $\bR_{k}$ can also describe the antenna correlation due to either insufficient antenna spacing or a lack of scattering.However, the following results focus only on diagonal $\bR_{k}$ matrices describing only large-scale fading, and $\bw_{k,n} \in \bbC^{M}$ is an uncorrelated fast-fading Gaussian channel vector drawn as  a realization from a zero-mean circularly symmetric complex Gaussian distribution, i.e., $\bw_{k,n} \sim \cC\cN(\b0,\bI_{M})$.}. 

\subsection{Phase Noise Model}
In practice, both  the transmitter and the receiver are impaired by phase noise induced during the up-conversion of the baseband signal to passband and vice-versa\footnote{The conversion takes place by multiplying the signal with the LO's output.}. As a result, the received signal is distorted during the reception processing, while at the transmitter side, a mismatch appears between the signal that  is intended to be transmitted and the generated signal.

In our analysis regarding the $n$th time slot, we take into account for non-synchronous operation at the BS, i.e., the BS antennas have independent phase noise processes $\phi_{m,n}, m=1,\ldots,M$ with $\phi_{m,n}$ being the phase noise process at the $m$th antenna. Note that the  phase noise processes are considered as mutually independent, since each antenna has its own oscillator, i.e., separate LOs (SLOs) at each antenna~\cite{Pitarokoilis2015,Krishnan2015}.  In case that all the antennas have a common LO (CLO), the phase noise processes $\phi_{m,n}$ are identical for all $m = 1, \ldots, M$ (synchronous operation). Similarly, we denote $\varphi_{k,n}$, $k=1,\ldots, K$ the phase noise process at the single-antenna user $k$.

Phase noise during the $n$th symbol can be described by a discrete-time  independent Wiener process, i.e., the phase noises at the  LOs of the $m$th antenna of the BS and $k$th user are modeled as
\begin{align}
 \phi_{m,n}&=\phi_{m,n-1}+\delta^{\phi_{m}}_{n}\label{phaseNoiseBS}\\
 \varphi_{k,n}&=\varphi_{k,n-1}+\delta^{\varphi_{k}}_{n},\label{phaseNoiseuser}
\end{align}
where $\delta^{\phi_{m}}_{n}\sim \cN(0,\sigma_{\phi_{m}}^{2}) $ and $\delta^{\varphi_{k}}_{n}\sim \cN(0,\sigma_{\varphi_{k}}^{2})$~\cite{Demir2000,Pitarokoilis2015,Krishnan2015}. Note that $\sigma_{i}^{2}=4\pi^{2}f_{\mathrm{c}} c_{i}T_{\mathrm{s}}$, $i=\phi_{m}, \varphi_{k}$ describes the phase noise increment variance with $T_{\mathrm{s}}$, $c_{{i}}$, and $f_{\mathrm{c}}$ being the  symbol interval, a constant dependent on the oscillator, and the carrier frequency, respectively. 

The phase noise, being the distortion in the phase due to the random phase drift  in the signal coming from the LOs of the BS and user $k$,  is expressed as a multiplicative factor to the channel vector as 
\begin{align}
\bg_{k,n}=\bTheta_{k,n}\bh_{k,n},
\end{align}
where $\bTheta_{k,n}\!\triangleq\!\mathrm{diag}\!\left\{ e^{j \theta_{k,n}^{(1)}}, \ldots, e^{j \theta_{k,n}^{(M)}} \!\right\}=e^{j \varphi_{k,n}}\bPhi_{n}\in \mathbb{C}^{M\times M}$ with $\bPhi_{n}\!\triangleq\!\mathrm{diag}\!\left\{\! e^{j \phi_{1,n}}, \ldots, e^{j \phi_{M,n}} \!\right\}$ being the phase noise sample matrix  at time $n$ because of the imperfections in the LOs of the BS,  while $e^{j \varphi_{k,n}}$ corresponds to the phase noise contribution by the LO of user $k$ \cite{Pitarokoilis2015,Krishnan2015}.
\subsection{Channel Estimation}
In realistic conditions,  where no perfect CSI is available, the BS needs to estimate the uplink channel~\cite{Papazafeiropoulos2015a,Ngo2013,Hoydis2013}. We take advantage of the TDD operation scheme, where  an uplink training phase during each transmission block (coherence period) exists by means of transmit pilot symbols. Since in TDD the number of required pilots scales with the number of terminals $K$, but not the number of BS antennas $M$, not only the analysis of systems with large number of BS antennas is  possible because of the finite duration of the coherence time\footnote{TDD is the only viable solution in fast time-varying channel conditions with massive MIMO systems, where the limitations for the coherence time are more stringent, while FDD can be considered only when there is some kind of channel sparsity~\cite{Bjoernson2015}.}, but also the knowledge of the downlink channel is considered known due to the property of reciprocity.

By assuming that the channel estimation takes place at time $0$, we now derive the joint channel-phase noise linear minimum mean-square
error detector  (LMMSE) estimate of the effective channel $\bg_{k,0}=\bTheta_{k,0}\bh_{k,0}$ in the presence of phase noise, small-scale fading, and channel impairments such as path loss. During this phase, we neglect the channel aging effect by assuming that both the channel and phase noise remain constant~\cite{Truong2014}. This is a valid assumption since the duration of the training phase is small, and the consequent time-variation of the channel is unnoticeable.  However, in the data transmission phase, taking place for ($T_{\mathrm{c}} -\tau$ ) symbols, the channel is supposed to vary from symbol to symbol, where $\tau$ and $T_{\mathrm{c}}$ are  the duration of the  training sequence during the training phase and   the channel coherence time, respectively.
\begin{proposition}\label{LMMSE} 
The LMMSE estimator of $\bg_{k,0}$, obtained during the training phase, is
 \begin{align}\label{estimatedChannel}
 \hat{\bg}_{k,0}=\left(  \Id_{M}+\frac{\sigma_{\mathrm{b}}^{2}}{p_{\rp}}\bR_{k}^{-1}\right)^{-1}\tilde{\by}_{k,0}^{\rp},
\end{align}
where $\sigma_{\mathrm{b}}^{2}$ is the variance of the post-processed noise at
base station, $\tilde{\by}_{,k,0}^{\rp}$ is a noisy observation of the effective channel from user $k$ to the BS, and $p_{\mathrm{p}}=\tau p_{\mathrm{u}}$  with $p_{\mathrm{u}}$   being the power per user in the uplink data transmission phase.
\end{proposition}
\proof: The proof of Proposition~\ref{LMMSE} is given in Appendix~\ref{proposition1}.\endproof

\subsection{Channel Aging}
The relative movement between the users and the BS inflicts a phenomenon, known as channel aging in the literature~\cite{Truong2013,Papazafeiropoulos2014,Papazafeiropoulos2014WCNC}. Specifically, the  Doppler shift arising from this movement causes the channel to vary between when it is learned by estimation, and when the estimate is applied for detection/precoding. Especially, channel aging is more severe in massive MIMO systems, where  narrow transmit beams appear due to the high angular resolution. Thus, the precoders that are used for achieving this spatial focus, bring a critical performance loss, which worsens as the users move with increasing velocity~\cite{Ying2014}. An autoregressive model of order $1$ enables  the  joint channel-phase noise process $\bg_{k,n}$ between the BS and the $k$th user at time $n$ (current time) to be approximated as\footnote{Contrary to other works~\cite{Vu2007,Truong2013,Papazafeiropoulos2014,Papazafeiropoulos2014WCNC}, $\bA_{n}$, being the factor of $\bg_{k,0}$,   is described by a matrix  instead of a scalar to account for the setting of SLOs. Moreover, in this work, $\bA_{n}$ is time-dependent by the  delay  $n$ in terms of the number of symbols transmitted, in order to include the accumulative channel aging effect~\cite{Kong2015a}.  Therefore, the longer the delay, the smaller the $\bA_{n}$, which in turn results in less accurate CSI. Hence, the channel estimated at time $0$, is used at time $1$, $2$, and so on till $n$. It is intuitive that the CSI will be much less reliable at time $n$, which results to a more realistic model. We assume that
it can be embodied in the channel vector $\bg_{k,0}$ by employing the circular symmetry of the channel distribution, or mainly, by the inclusion of an appropriate scaling of the covariance matrix.}
\begin{align}
\bg_{k,n}  = \bA_{n} \bg_{k,0} + \bee_{k,n},\label{eq:GaussMarkoModel}
\end{align}
where, initially, $\bA_{n}=\mathrm{J}_{0}(2 \pi f_{\mathrm{D}}T_{\mathrm{s}}n) \Id_{M}$ is supposed to model 2-D isotropic scattering~\cite{WCJr1974}\footnote{Normally, the BS antennas are deployed in a fixed space instead of considering  distantly distributed small BSs constituting one massive MIMO BS. In such case, $\bA_{n}$ is a scalar, since it can be  reasonably assumed that all the BS antennas have the same relative movement comparing to the user.}, $\bg_{k,0}$ is the channel vector received during the training phase, and $\bee_{k,n} \in \bbC^{M}$ is the uncorrelated channel error vector due to the channel variation modelled as a stationary Gaussian random process with independent and identically distributed (i.i.d.)~entries and distribution $\cC\cN(\b0,\bR_{k}-\bA_{n}\bR_{k}\bA_{n})$. 

This model can be enriched by incorporating other interesting effects. By introducing the effect of channel estimation to the autoregressive model of~\eqref{eq:GaussMarkoModel}, the BS  takes into account for both estimated and delayed CSI, in order to   design the detector $\bW_{n}\in \mathbb{C}^{M\times M}$ or the precoder $\bF_{n}\in \mathbb{C}^{M\times M}$ at time instance $n$ for the uplink and downlink, respectively. Hence, the effective channel at time $n$ is expressed by
\begin{align}
\bg_{k,n} 
= & \bA_{n} \bg_{k,0} + \bee_{k,n}\\
= & \bA_{n}\hat{\bg}_{k,0} + \tilde{\bee}_{k,n},\label{eq:GaussMarkov2}\end{align} 
where $\hat{\bg}_{k,0}\sim \cC\cN\left(\b0,\bD_{k}\right)$ with $\bD_{k}=\left(  \Id_{M}+\frac{\sigma_{b}^{2}}{p_{\rp}}\bR_{k}^{-1}\right)^{-1}\bR_{k}$ and  $\tilde{\bee}_{k,n}\triangleq \bA_{n} \tilde{\bg}_{k,n} +  \bee_{k,n}\sim \cC\cN(\b0, \bR_{k} - \bA_{n}\bD_{k}\bA_{n})$  are mutually independent, while $\bA_{n}$, is assumed to be known to the BS, now incorporates both the effects of 2-D isotropic scattering, user mobility,  and phase noise, as shown below by Theorem~\ref{LMMSE}. In other words, the combined error $\tilde{\bee}_{k,n}$ depends on both the imperfect and the general delayed CSI effects, allowing the elicitation of interesting outcomes during the ensuing analysis in Section~\ref{Deterministic}.  
  
We turn our attention to the derivation of $\bA_{n}$ by accounting for user mobility, phase noise, and the general channel effects modelled by  $\bR_{k}$.
\begin{Theorem}\label{LMMSE}
 The  effective joint channel-phase noise  vector at time $n$  is expressed as
 \begin{align}
  {\bg}_{k,n}^{\mathrm{Joint}}=\bA_{n} {\bg}_{k,0}
 \end{align}
with 
\begin{align}
\bA_{n}=\mathrm{J}_{0}(2 \pi f_{\mathrm{D}}T_{\mathrm{s}}n)e^{-\frac{\sigma_{\varphi_{k}}^{2}}{2}n}\Delta\bPhi_{n}, \label{agingParameter}
\end{align}
where $\Delta\bPhi_{n}=\mathrm{diag}\left\{e^{-\frac{\sigma_{\phi_{1}}^{2}}{2}n} ,\ldots,e^{-\frac{\sigma_{\phi_{M}}^{2}}{2}n}\right\}$.
 \end{Theorem}
\proof
The total mean square error (MSE), which is actually  given by means of the error covariance matrix, is $\mathrm{MSE}=\tr {\bE}_{k,n}$, where ${\bE}_{k,n}$ is the error covariance written as
\begin{align}
&\!\!{\bE}_{k,n}=\mathbb{E}\left[ \left( \bg_{k,n}-\bA_{n} {\bg}_{k,0} \right) \left( \bg_{k,n}-\bA_{n} {\bg}_{k,0} \right)^{\H}\right]\nn\\
&\!\!=\mathbb{E}| \bg_{k,n}|^{2}+\bA_{n} \mathbb{E}|{\bg}_{k,0}|^{2}\bA_{n}^{\H}-2\mathbb{E}\left[ \mathrm{Re}\left\{\bA_{n}  {\bg}_{k,0}\bg_{k,n}^{\H} \right\} \right]\nn\\
&\!\!=\bR_{k}+\bA_{n}\bR_{k}\bA_{n}^{\H}-2\mathbb{E}\left[ \mathrm{Re}\left\{\bA_{n}  {\bg}_{k,0}\bg_{k,n}^{\H} \right\} \right]\nn\\
&\!\!\!\!=\bR_{k}\!+\!\bA_{n}\bR_{k}\bA_{n}^{\H}\!-\!2 {\mathrm{J}_{0}}(2 \pi f_{\mathrm{D}}T_{\mathrm{s}}n)e^{-\frac{\sigma_{\varphi_{k}}^{2}}{2}n}\!\bA_{n} \bR_{k}\Delta\bPhi_{n}.\label{error_Cov_mat}
\end{align}
Note that the expectation is taken over all channel and phase noise  realizations. Also,  we have used that $\bg_{k,n}=\bTheta_{k,n}\bh_{k,n}$ with $\bTheta_{k,n}$ and $\bh_{k,n}$ being uncorrelated. Also, we have denoted $\Delta\bPhi_{n}=\mathrm{diag}\left\{e^{-\frac{\sigma_{\phi_{1}}^{2}}{2}n} ,\ldots,e^{-\frac{\sigma_{\phi_{M}}^{2}}{2}n}\right\}$, and  
\begin{align}
\EE [{\bh}_{k,0}{\bh}_{k,n}^{\H}] = \mathrm{J}_{0}(2 \pi f_{D}T_{s}n)\bR_{k}.\label{eq:autoCorrelation2}
\end{align}
The multiplcative parameter matrix $\bA_{n}$ is determined by minimizing the MSE in~\eqref{error_Cov_mat}. Thus, if we differentiate~\eqref{error_Cov_mat} with respect to $\bA_{n}$, and equate the resulting expression to zero,  we obtain the Hermitian  matrix $\bA_{n}$ as in~\eqref{agingParameter}.
\endproof
\begin{remark}
 During the data transmission phase, the available CSI at time $n$ is obtained in terms of the estimated CSI at time 0. The parameter matrix $\bA_{n}$  has a physical meaning, since it describes the effects of 2-D isotropic scattering, user mobility, and phase noise impairments at  both the UE and the BS. 
\end{remark}
\begin{remark}
Notably, phase noise induces an extra loss due to the additional deviation between the actual channel at time n, and the channel estimate from time 0 because of the symbol-by-symbol random phase drift. In other words, the phase noise contributes to the channel aging phenomenon that  imposes a further challenge to  investigate the realistic potentials of massive MIMO systems.
\end{remark}

\begin{corollary}
 In the case where  all the BS antennas are connected to the same oscillator or  all the oscillators are identical, $\bA_{n}$ degenerates to a scalar $\al_{n}= \mathrm{J}_{0}(2 \pi f_{\mathrm{D}}T_{\mathrm{s}}n)e^{-\frac{\sigma_{\varphi_{k}}^{2}+\sigma_{\phi_{k}}^{2}}{2}n}$ or a scaled identity matrix $\al_{n}\Id_{M}$, respectively. 
\end{corollary}

\section{Downlink Transmission}
The BS transmits data simultaneously to all users in its cell by employing space-division multiple access (SDMA). Applying reciprocity, the downlink channel is written as the Hermitian  transpose of the uplink channel. The coherence time,  the received signal $y_{k,n}\in\bbC$ by the $k$th UE during the data transmission phase ($n=\tau+1,\ldots, T_{\mathrm{c}}$) is
\begin{align}
\!\!y_{k,n}=\sqrt{p_{\mathrm{d}}}\bh^\H_{k,n}\bTheta_{k,n}\bs_{n}+z_{k,n}\label{eq:DLreceivedSignal}
\end{align}
where  $\bs_{n}=\sqrt{\lambda}\bF_{n}\bx_{n}$ denotes the signal vector transmitted by the BS with $\lambda$, $\bF_{n} \in \bbC^{M \times K}$ and  $\bx_{n} = \big[x_{1,n},~x_{2,n},\cdots,~x_{K,n}\big]^\T \in \bbC^{K}\sim \cC\cN(\b0,\bI_{K})$ being a normalization parameter, the linear precoding matrix,  and the data symbol vector to its $K$ served users, respectively. Moreover, we have assumed that the BS transmits data to all the users with the same power $p_{\mathrm{d}}$, and $z_{k,n} \sim \cC\cN(0,\sigma_{k}^{2})$ is complex Gaussian noise at user $k$.  The need for constraining the transmit power per user to $p_{\mathrm{d}}$, i.e., $\mathbb{E}\left[\frac{p_{\mathrm{d}}}{K}\bs_{n}^{\H}\bs_{n}\right]=p_{\mathrm{d}}$, provides the normalization parameter as
\begin{align}
\lambda=\frac{1}{\EE \left[\frac{1}{K} \tr\bF_{n}\bF^\H_{n}\right]}. \label{eq:lamda} 
\end{align}

Henceforth, by taking into account that $\bg_{k,n}=\bTheta_{k,n}\bh_{k,n}$, we proceed with the following presentation of the received signal, obtained by~\eqref{eq:DLreceivedSignal}.  Thus, we have
\begin{align}
&\!\!y_{k,n}\!\!=\!\!{\sqrt{\!\lambda p_{\mathrm{d}}}}\bg^\H_{k,n}\!\bTheta_{k,n}^{2} {\bff}_{k,n}x_{k,n}
\!\!+\!\!\sum_{i \neq k}\!\! {\sqrt{\!\lambda p_{\mathrm{d}}}}\bg^\H_{k,n}\!\bTheta_{k,n}^{2} { \bff}_{i,n}x_{i,n}\!+\!z_{k,n}\label{eq:DLreceivedSignal1}\\
\!\!&=\!\!\underbrace{{ \sqrt{\!\lambda p_{\mathrm{d}}}}\mathbb{E}\!\!\left[\bg^\H_{k,n} \!\bTheta_{k,n}^{2} {\bff}_{k,n}\right]\!x_{k,n}}_{\mathrm{desired~signal}}\!+\!\underbrace{z_{k,n}}_{\mathrm{noise}}
\!\!+\!\!\sum_{i \neq k} \!{\sqrt{\!\lambda p_{\mathrm{d}}}}\bg^\H_{k,n}\! \bTheta_{k,n}^{2} {\bff}_{i,n}x_{i,n}\nn\\
\!\!&+\!{ \sqrt{\!\lambda p_{\mathrm{d}}}}\left(\bg^\H_{k,n}\bTheta_{k,n}^{2} {\bff}_{k,n}x_{k,n}\!-\!\mathbb{E}\!\!\left[\bg^\H_{k,n} \bTheta_{k,n}^{2} {\bff}_{k,n}\right]\!x_{k,n}\right)\!.\label{eq:DLreceivedSignalgeneral}
\end{align}

In~\eqref{eq:DLreceivedSignal1}, we have  applied a similar technique to~\cite{Medard2000} because UEs do not have access to instantaneous CSI during the downlink phase, but we can  assume that, in particular, UE $k$  is aware of only $\EE[\bg^\H_{k,n} \bff_{jm} ]$.

Given that the input symbols are Gaussian, we provide the achievable user rate based on the analysis in~\cite[Theorem $1$]{Hassibi2003}. Specifically, the worst case uncorrelated additive noise is also zero-mean with variance equal to the variance of interference plus noise. In addition, since $\bg_{k,n}$ is circularly symmetric, we let $\bg_{k,n}=\bTheta_{k,0}^{2}\bg_{k,n}$. Thus, by denoting $\widetilde{\bTheta}_{k,n}=\Delta{\bTheta}_{k,n}$,  where $\Delta{\bTheta}_{k,n}=\left( \bTheta_{k,n}{\bTheta_{k,0}}^{\!\!\!\H} \right)^{2}=\mathrm{diag}\left\{ e^{2 j \left(\theta_{k,n}^{(1)}-\theta_{k,0}^{(1)}  \right)}, \ldots, e^{2 j \left(\theta_{k,n}^{(1)}-\theta_{k,0}^{(1)}  \right)} \right\}$, we may consider a SISO model with the desired signal power and the interference plus noise power at UE $k$ given by 
 \begin{align}
S_{k,n}
=  \lambda\Big|\EE\left[\bg^\H_{k,n}\widetilde{\bTheta}_{k,n}{\bff}_{k,n}\right]  \Big|^{2},\label{eq:DLgenSignalPower}
\end{align}
and 
\begin{align}
I_{k,n}
&\!=\!   \lambda\var\!\left[\bg^\H_{k,n}\widetilde{\bTheta}_{k,n} \bff_{k,n}\right] \! +  \! \frac{\sigma_{k}^{2}}{p_{\mathrm{d}}} \!+\!    \sum_{i\ne k}   \!  \lambda\EE\bigg[\Big|\bg^\H_{k,n}\widetilde{\bTheta}_{k,n} \bff_{i,n}\Big|^{2}\bigg] .   \label{eq:DLgenIntfPower}
\end{align}

Since the available CSI at the BS at time $n$ is $\bA_{n}\hat{\bg}_{k,0}$,   as can be seen from~\eqref{eq:GaussMarkov2}, the expression corresponding to the MRT precoder is
\begin{align}
\bF_{n}&=  \bA_{n}{\hat{\bG}}_{0}. \label{eq:precoderMRT}
\end{align}

In a similar way, 
the BS  designs its RZF precoder as\! \cite{Hoydis2013}
\begin{align}
\bF_{n}
&= \left(\bA_{n}{\hat{\bG}}_{0}{\hat{\bG}}_{0}^\H \bA_{n} \!+\! \bZ + M  a \Id_M\right)^{-1}\bA_{n}{\hat{\bG}}_{0}\nn\\
&= {\bSigma} \bA_{n}{\hat{\bG}}_{0}, \label{eq:precoderRZF}
\end{align}
where we define ${\bSigma}\triangleq\left({\bA_{n}{\hat{\bG}}_{0}{\hat{\bG}}_{0}^{\H}\bA_{n}} + \bZ +   a M \Id_M\right)^{-1}$ with $\bZ \in \bbC^{M \times M}$ being an arbitrary Hermitian
nonnegative definite matrix and $a$ being a regularization scaled by $M$, in order to converge to a constant, as $M$, $K\to \infty$. Note that both $\bZ$ and $a$ could be optimized, but this is outside the
scope of this paper and left to future work. One possible common choise for these parameters is provided in~\cite{Hoydis2013}.

Following a similar approach to~\cite{Bjornson2015,Pitarokoilis2015}, we compute the achievable rate for each time instance of the data transmission phase. Thus, the downlink ergodic achievable rate of user $k$, is given by
\begin{align}
R_{k}& = \frac{1}{T_{c}}\sum_{n=1}^{T_{c}-\tau}R_{k,n}\nn\\
 &=\frac{1}{T_{c}}\sum_{n=1}^{T_{c}-\tau}\mathrm{log}_{2}\left(1+ \gamma_{k,n} \right),
\label{eq:DLrate}
\end{align}
where $\gamma_{k,n}=\frac{S_{k,n} }{I_{k,n}}$ is the instantaneous downlink SINR at time $n$.  In other words, the mutual information between the received signal and the transmitted symbols is lower bounded by the achievable rate per user provided in~\eqref{eq:DLrate}. 

\section{Deterministic Equivalent Analysis}\label{Deterministic}
In this section, we provide the derivations of the downlink SINRs after applying  MRT and RZF precoders in the presence of imperfect  and delayed CSI due to user mobility and phase noise. Use of RMT in terms of DEs enables the extraction of asymptotic expressions as $K,M \rightarrow \infty$, while keeping their ratio ${K}/{M}=\beta$ finite. Given that the theory of  DEs provides tight approximations  even for moderate system dimensions, the significance of our  results is of great importance. Confirmation of this tightness is provided in Section~\ref{results} by  simulations. Interestingly, the results mirror the dependence of the harmful effects described by channel aging.

The DE of the SINR $\gamma_{k,n}$ is such that $\gamma_{k,n}-\bar{\gamma}_{k,n}\xrightarrow[M \rightarrow \infty]{\mbox{a.s.}}0$\footnote{Note that $\xrightarrow[ M \rightarrow \infty]{\mbox{a.s.}}$ denotes almost sure convergence, and  $a_n\asymp b_n$ expresses the equivalence relation $a_n - b_n  \xrightarrow[ M \rightarrow \infty]{\mbox{a.s.}}  0$ with $a_n$  and $b_n$  being two infinite sequences.}, while the deterministic rate of user $k$ is obtained by the dominated convergence~\cite{Billingsley2008} and the continuous mapping theorem~\cite{Vaart2000} by means of~\eqref{eq:DLrate} 
\begin{align}
R_{k}-\frac{1}{T_{\mathrm{c}}}\sum_{n=1}^{T_{\mathrm{c}}-\tau}\log_{2}(1 + \bar{\gamma}_{k,n}) \xrightarrow[ N \rightarrow \infty]{\mbox{a.s.}}0.\label{DeterministicSumrate}
\end{align}

The DE downlink achievable  user rates corresponding to MRT and RZF can be obtained by means of the following theorems. Hereafter, for the sake of exposition and without loss of generality, we assume that in the case of  SLOs, the phase noises obey to identical statistics. However, in Appendix~\ref{theorem2}, we scrutinize the general setting, where  phase noises across different LOs  are independent, but not identically distributed. First, we consider MRT precoding. 
\subsection{MRT}
This section presents the general DE regarding the SINR with MRT precoding. Then, this asymptotic result particularizes to a number of cases of interest. In additions, it is investigated the dependence of the transmit power per user with the number of BS antennas (power scaling law), in order to support a specific desired rate.
\begin{Theorem}\label{theorem:DLagedCSIMRT}
The downlink DE of the SINR of user $k$ at time $n$ with MRT precoding , accounting for imperfect CSI and  delayed CSI due to phase noise and user mobility, is given by 
\begin{align}
\bar{\gamma}_{k,n} = \frac{  e^{-2\left( \sigma_{\varphi_{k}}^{2}+\sigma_{\phi}^{2} \right)n} {\delta}_{k}^{2}}{{\frac{1}{M}{\delta}_{k}^{'}}+  \frac{\sigma_{k}^{2}}{p_{\mathrm{d}}  \bar{\lambda}M}
 + \sum_{i\neq k}\frac{1}{M}\delta_{i}^{''}},\label{eq:DLdelayedCSIetaMRT1}
 \end{align}
 with 
  \begin{align}
  \bar{\lambda}&=\left( \frac1K\sum_{i=1}^{K}\frac{1}{M}\tr \bA_{n}^{2}\bD_{k} \right)^{-1},\nn
  \end{align}
${\delta}_{k}=\frac{1}{M}\tr\bA_{n}^{2}\bD_{k}$, ${\delta}_{k}^{'}=\frac{1}{M}\tr\bA_{n}^{2} \bD_{k} \left( \bR_{k} - \bA_{n}^{2}\bD_{k} \right)$, and $\delta_{i}^{''}=\frac{1}{M} \tr\bA_{n}^{2}\bD_{i} \bR_{k}$.
\end{Theorem}
\proof: The proof of Theorem~\ref{theorem:DLagedCSIMRT} is given in Appendix~\ref{theorem2}.\endproof
\begin{corollary}\label{NoCorrelation}
In  case that $\bR_{k}=\Id_{M}$ (no large-scale component), and $\bA_{n}$ degenerates to a scalar $\al_{n}$ or a $\al_{n}\Id_{M}$, i.e., the BS employs a CLO or a SLOs setting with identical LOs, respectively, we obtain
 \begin{align}
  \bar{\gamma}_{k,n} =\frac{\al_{n}^{2} d  e^{-2\left( \sigma_{\varphi_{k}}^{2}+\sigma_{\phi}^{2} \right)n} }{  \frac{1-\al_{n}^{2} d}{M}+  \frac{\sigma_{k}^{2}}{p_{\mathrm{d}} M}
 +{  \frac{ K-1 }{M} }},\label{eq:DLdelayedCSIetaMRT3}
 \end{align}
 where $d=\left( \frac{\tau p_{ \mathrm{u}}}{\tau p_{\mathrm{u}}+\sigma_{b}^{2}}\right)^{-1}$ and $\al_{n}= \mathrm{J}_{0}(2 \pi f_{\mathrm{D}}T_{\mathrm{s}}n)e^{-\frac{\sigma_{\varphi_{k}}^{2}+\sigma_{\phi}^{2}}{2}n}$.
 \end{corollary}
\proof Using $\bR_{k}=\Id_{M}$ leads to $\bD_{k}=d\,\Id_{M}=\left( \frac{\tau p_{ \mathrm{u}}}{\tau p_{\mathrm{u}}+\sigma_{b}^{2}}\right)^{-1}\Id_{M}$. Moreover, if $\bA_{n}$ is replaced by the scalar $\al_{n}$ or the scaled identity matrix $\al_{n}\Id_M$,   Theorem~\ref{theorem:DLagedCSIMRT} allows to simplify enough the deterministic SINR. Hence, by substituting  $\bar{\lambda}=\left(\al_{n}^{2} d\right)^{-1}$, ${\delta}=\al_{n}^{2} d$, ${\delta}^{'}=\al_{n}^{2} d \left( 1 - \al_{n}^{2} d \right)$, and $\delta^{''}=\al_{n}^{2} d$, we obtain $\bar{\gamma}_{k,n}$ as in~\eqref{eq:DLdelayedCSIetaMRT3}.
\endproof

Nevertheless, it is interesting to investigate the effect of the combined impairments of phase noise and user mobility  on the asymptotic power scaling law. For the sake of exposition, we focus on MRT precoding, however similar results can be obtained in the case of RZF precoding as well.
\begin{proposition}\label{powerLaw}
 For the identical SLOs or CLO setting, the downlink achievable SINR of user $k$ at time $n$ with MRT precoding  subject to imperfect CSI, delayed CSI due to phase noise  and user mobility, and collocated  BS antennas  becomes
 \begin{align}
  \gamma_{k,n}=\frac{\tau E_{\mathrm{d}}E_{\mathrm{u}}}{\sigma^{4}M^{2 q-1}}\mathrm{J}_{0}^{2}(2 \pi f_{\mathrm{D}}T_{\mathrm{s}}n)e^{-3\left( {\sigma_{\varphi_{k}}^{2}+\sigma_{\phi}^{2}} \right)n}[\bR_{k}^{2}]_{mm},\label{scalingPower}
 \end{align}
where the transmit uplink and downlink powers are scaled proportionally to $1/M^{q}$, i.e.,  $p_{\mathrm{u}} = E_{\mathrm{u}}/M^{q}$ and $p_{\mathrm{d}}= E_{\mathrm{d}}/M^{q}$
for fixed $E_{\mathrm{u}}$ and $E_{\mathrm{d}}$, and $q>0$.
\end{proposition}
\proof
Let us first substitute the MRT precoder, given by~\eqref{eq:precoderMRT},  in~\eqref{eq:DLreceivedSignalgeneral}. Then, we divide both the desired and the interference parts by $\lambda$ and $1/M^{2}$. The desired signal power is written as
\begin{align}
S_{k,n}^{\mathrm{MRT}}
&=  \frac{1}{M^{2}}\Big|\bg^\H_{k,n} \widetilde{\bTheta}_{k,n} \bA_{n} {\hat{\bg}}_{k,0}\Big|^{2}\nn\\
&=  \frac{1}{M^{2}}\Big|\hat{\bg}^\H_{k,0}\widetilde{\bTheta}_{k,n}  \bA_{n}^{2} {\hat{\bg}}_{k,0}\Big|^{2}\label{desiredMid}\\
&=\frac{1}{M^{2}}\mathrm{J}_{0}^4{2}(2 \pi f_{\mathrm{D}}T_{\mathrm{s}}n)e^{-4 \left( {\sigma_{\varphi_{k}}^{2}+\sigma_{\phi}^{2}} \right)n}[\bD_{k}]_{mm}^{2},\label{desired}
\end{align}
where $[\bD_{k}]_{mm}$ is the $m$th diagonal element of the matrix  $\bD_{k}$ expressing the variance of the $m$th element. Given that both $\widetilde{\bTheta}_{k,n}$ and $  \bA_{n}$ are diagonal matrices,   we have taken into account in~\eqref{desired} that they can commute. Moreover, in the last step of~\eqref{desired}, we have used the law of large numbers for large $M$ in the case of collocated  BS antennas, as well as that $\bA_{n}$ is a scaled identity matrix  or a scalar in the case of identical SLOs or CLO, respectively. In other words, $ {\hat{\bg}}_{k,0}$ has i.i.d. elements with variance $[\bD_{k}]_{mm}$. As far as the interference is concerned, the first and third terms of~\eqref{eq:DLgenIntfPower} vanish to zero as $M \to \infty$ by means of the same law. Thus, we have
\begin{align}
I_{k,n}^{\mathrm{MRT}}&=\frac{\sigma^{2}_{k}}{M^{2}p_{\mathrm{d}} \lambda}\nn\\
&=\frac{\sigma^{2}_{k}\mathrm{J}_{0}^{2}(2 \pi f_{\mathrm{D}}T_{\mathrm{s}}n)e^{-\left( {\sigma_{\varphi_{k}}^{2}+\sigma_{\phi}^{2}} \right)n}[\bD_{k}]_{mm}}{M p_{\mathrm{d}} }, 
\end{align}
where  we have applied the law of large numbers to obtain $ \lambda\!=\!\left(\frac{1}{M}\mathrm{J}_{0}^{2}(2 \pi f_{\mathrm{D}}T_{\mathrm{s}}n)e^{-\left( {\sigma_{\varphi_{k}}^{2}\!+\sigma_{\phi}^{2}} \right)n}[\bD_{k}]_{mm}\! \right)^{-1}\!$ from~\eqref{eq:lamda}, and we have substituted  $\bD_{k}=\frac{\tau E_{\mathrm{u}}}{M^{q}\sigma^{2}_{k}}\bR_{k}^{2}$, since  $p_{\mathrm{p}}$ depends on $p_{\mathrm{u}}$, the result in~\eqref{scalingPower} is obtained. \endproof

Proposition~\ref{powerLaw} reveals that the selection of $q$ affects heavily the achievable rate per user. In fact, proper selection allows maintaining the same performance, even by further scaling down  the transmit power of each user. Specifically, if $q$ is set less than $1/2$, $\gamma_{k,n}$ is unbounded, i.e., it tends to infinity.  On the contrary, the SINR of user $k$ diminishes to zero, if $q>1/2$. This clearly indicates that the transmit powers of each user during the training and downlink phases have been reduced over the required value. Most importantly, in the special case that $q=1/2$, the SINR approaches a non-zero limit given by the following corollary.
\begin{corollary}
In the presence of phase noise and user mobility, when the transmit uplink and downlink powers are scaled down by $p_{\mathrm{u}} = E_{\mathrm{u}}/\sqrt{M}$ and $p_{\mathrm{d}} = E_{\mathrm{d}}/\sqrt{M}$ for fixed $E_{\mathrm{u}}$ and $E_{\mathrm{d}}$, the downlink SINR per user with MRT can be finite  as
\begin{align}
   \gamma_{k,n}=\frac{\tau^{2} E_{\mathrm{d}} E_{\mathrm{u}}}{\sigma_{k}^{4}}\mathrm{J}_{0}^{2}(2 \pi f_{\mathrm{D}}T_{\mathrm{s}}n)e^{-3\left( {\sigma_{\varphi_{k}}^{2}+\sigma_{\phi}^{2}} \right)n}[\bR_{k}^{2}]_{mm}.
\end{align}
\end{corollary}

Evidently, phase noise and user mobility reduce the SINR, but the power scaling law is not affected. 
\subsection{RZF}
As far as RZF is concerned, the analysis is more complex, as shown below. First we present the DE of the SINR in general channel aging conditions by means of the following theorem.
\begin{Theorem}\label{theorem:DLagedCSIRZF}
The downlink DE of the SINR user $k$ at time $n$ with RZF precoding, accounting for imperfect CSI and delayed CSI due to phase noise and user mobility, is given by 
\begin{align}
\bar{\gamma}_{k,n} = \frac{  e^{-2\left( \sigma_{\varphi_{k}}^{2}+\sigma_{\phi}^{2} \right)n}  {\delta}_{k}^{2}}{{\frac{1}{M}{\delta}_{k}^{'}}+  \frac{\sigma_{k}^{2}\left( 1+\delta_{k} \right)^{2}}{p_{\mathrm{d}}  \bar{\lambda}}
 + \sum_{i\neq k}{ \frac{{Q}_{ik}\left( 1+\delta_{k} \right)^{2}}{M\left(1+{\delta_{i}}\right)^{2}} }},\label{eq:DLdelayedCSIetaRZF}
 \end{align}
 with 
  \begin{align}
  \bar{\lambda}&=\frac{K}{ \left(\frac{1}{M}\tr\Tm - \frac{1}{M} \tr \left(\frac{\Zm}{M} +a \Id_M\right) {\bC}\right)}\nn\\
{Q}_{ik}&\asymp \frac{1}{M}\tr\bA_{n}^{2}\bD_{i}\bC^{'''}
+\frac{\left|{\delta_{k}}\right|^{2}\delta_{k}^{''}}{\left( 1+\delta_{k} \right)^{2}}-2\mathrm{Re}\left\{  \frac{\delta_{k}\delta_{k}^{''} }{\left( 1+\delta_{k} \right)}\right\},\nn
  \end{align}
  ${\delta}_{k}=\frac{1}{M}\tr\bA_{n}^{2}\bD_{k}\bT,
 {\delta}_{k}^{'}=\frac{1}{M}\tr\bA_{n}^{2}\bD_{k}{\bC}^{'}, \mathrm{and}~
 \delta_{k}^{''}=\frac{1}{M}\tr\bA_{n}^{2}\bD_{k}\bC^{''}$,
 where
\begin{itemize}
\renewcommand{\labelitemi}{$\ast$}
\item $\bT=\bT(a)$ and ${\deltav}=[{\delta}_{1},\cdots,{\delta}_{K}]^\T={\deltav}(a)={\ev}(a)$ are given by Theorem~\ref{th:detequ} for $\bL=\bA_{n}^{2}\bD_{k}$, $\bS=\bZ/M$, $\bR_k=\bA_{n}^{2}\bD_{k}\, \forall k \in \mathcal{K}$,
\item  $\bC=\bT^{'}(a)$ is given by Theorem~\ref{th:detequder} for $\bL=\Id_M$, $\bS=\bZ/M$, $\bK=\Id_M$, $\bR_k=\bA_{n}^{2}\bD_{k}\, \forall  k \in \mathcal{K}$,
\item  $\bC^{'}=\bT^{''}(a)$ is given by Theorem~\ref{th:detequder} for $\bL=\bA_{n}^{2}\bD_{k}$, $\bS=\bZ/M$, $\bK= \bR_{k} - \bA_{n}^{2}\bD_{k}$, $\bR_k=\bA_{n}^{2}\bD_{k}\, \forall  k \in \mathcal{K}$,
\item  $\bC^{''}=\bT^{'''}(a )$ is given by Theorem~\ref{th:detequder} for $\bL=\bA_{n}^{2}\bD_{k}$, $\bS=\bZ_j/M$, $\bK=\bA_{n}^{2}\bD_{i}$, $\bR_k=\bA_{n}^{2}\bD_{k}\, \forall  k \in \mathcal{K}.$
\item  $\bC^{'''}=\bT^{''''}(a)$ is given by Theorem~\ref{th:detequder} for $\bL=\bA_{n}^{2}\bD_{i}$, $\bS=\bZ_j/M$, $\bK=\bR_{k}$, $\bR_k=\bA_{n}^{2}\bD_{k}\, \forall  k \in \mathcal{K}.$
\end{itemize} 
\end{Theorem}
\proof: The proof of Theorem~\ref{theorem:DLagedCSIRZF} is given in Appendix~\ref{theorem3}.\endproof
\begin{corollary}\label{equal_correlation}
 Let $\bR_{k}=\bR$, i.e., the large-scale effects (path loss and shadowing) affect the same all users, then $\bD_{k}=\bD$. In such case, the deterministic SINR of Theorem~\ref{theorem:DLagedCSIRZF}  $\bar{\gamma}_{k,n}$ can be simplified to
 \begin{align}
  \bar{\gamma}_{k,n} = \frac{  e^{-2\left( \sigma_{\varphi_{k}}^{2}+\sigma_{\phi}^{2} \right)n}    {{\delta}}^{2}}{  {\frac{1}{M}{\delta}^{'}}+  \frac{\sigma_{k}^{2}\left( 1+\delta \right)^{2}}{p_{\mathrm{d}} \bar{\lambda}}
 + \left( K-1 \right){  \frac{{Q}}{M} }},\label{eq:DLdelayedCSIetaRZF2}
 \end{align}
with  ${\delta}=\frac{1}{M}\bA_{n}^{2}\bD\bT\triangleq e,
 {\delta} ^{'}=\frac{1}{M}\bA_{n}^{2}\bD{\bC}^{'},  \delta ^{''}=\frac{1}{M}\bA_{n}^{2}\bD\bC^{''}$,  $ \bT = ({\bA_{n}^{2}\bD }/{\beta\left( 1+\delta \right)}$  $  +\bZ/M + a\Id_M)^{-1}$, $\bar{\lambda}={K}/{ \left(\frac{1}{M}\tr\Tm - \frac{1}{M} \tr \left(\frac{\Zm}{M} +a \Id_M\right) {\bC}\right)}$, and
 \begin{align}
  {Q}\asymp \frac{1}{M}\tr\bR \bC^{''}+\frac{\left|{\delta }\right|^{2}\delta ^{''}}{\left( 1+\delta  \right)^{2}}-2\mathrm{Re}\left\{  \frac{\delta \delta ^{''} }{\left( 1+\delta  \right)}\right\},\nn
 \end{align}
while the general expression of $\bT^{'}$ is given by 
 \begin{align}
\bT^{'}=\bT\bK\bT+e^{'}_{\bK}/{\beta\left(1+\delta\right)^2}\bT{\bA_{n}^{2}\bD  }\bT,  \nn
 \end{align}
where  $e^{'}_{\bK}=\beta \left( 1+\delta \right)^{2}e_{111}^{\bK}/\left( \beta-e_{201}^{\bK} \right)$  with  $e_{ijm}^{\bK}=1/\left( 1+\delta \right)^{j+m}\frac1M \tr\left( \bA_{n}^{2}\bD \right)^{i}\bT\bK^{j}\bT^{m}$.
 \end{corollary}
 \begin{corollary}\label{NoCorrelation1}
If no large-scale component $(\bR_{k}=\Id_{M}$) are assumed, and the phase noise  from all the oscillators is considered identical or the BS has only a CLO, i.e., $\bA_{n}$ becomes $\al_{n}\Id_M$ or a scalar $\al_{n}$, we obtain
 \begin{align}
  \bar{\gamma}_{k,n} = \frac{    e^{-2\left( \sigma_{\varphi_{k}}^{2}+\sigma_{\phi}^{2} \right)n}  {{\delta}}^{2}}{  {\frac{1}{M}{\delta}^{'}}+  \frac{\sigma_{k}^{2}\left( 1+\delta \right)^{2}}{p_{\mathrm{d}} \bar{\lambda}}
 + \left( K-1 \right){  \frac{{Q}}{M} }},\label{eq:DLdelayedCSIetaRZF3}
 \end{align}
 where $d=\left( \frac{\tau p_{ \mathrm{u}}}{\tau p_{\mathrm{u}}+\sigma_{b}^{2}}\right)^{-1}$ and $\al_{n}= \mathrm{J}_{0}(2 \pi f_{\mathrm{D}}T_{\mathrm{s}}n)e^{-\frac{\sigma_{\varphi_{k}}^{2}+\sigma_{\phi}^{2}}{2}n}$.
 \end{corollary}
\proof When $\bR_{k}=\Id_{M}$, $\bD_{k}$  becomes a scaled identity matrix, i.e., $\bD_{k}=d \Id_{M}=\left( \frac{p_{\rp}\tau}{p_{\rp}\tau+\sigma_{b}^{2}}\right)\Id_{M}$. In addition, replacing $\bA_{n}$ by the scalar $\al_{n}=\mathrm{J}_{0}(2 \pi f_{\mathrm{D}}T_{\mathrm{s}}n)e^{-\frac{\sigma_{\varphi_{k}}^{2}+\sigma_{\phi_{k}}^{2}}{2}n}$, we obtain $\delta=\al_{n}^{2} d  t$, 
 ${\delta} ^{'}=\al_{n}^{2}dt^{'}_1$,  $\delta ^{''}=\al_{n}^{2} d t^{'}_{2}$,   $\bar{\lambda}=\frac{K}{ \left(t -t^{'}_{3}a- \frac{ t^{'}_{3}z}{M}  \right)}$, 
 \begin{align}
  {Q}&\asymp\frac{\beta d^4 \delta^2 \left(\delta^2+\delta+1\right)}{\beta (\delta+1)^2-d^4 \delta^2}.\nn
  \end{align}
 By considering the unique positive root of the quadratic equation  $\delta=\al_{n}^{2}d t$, we obtain~\eqref{delta}.
 \begin{figure*}
 \begin{align}
 \delta =\frac{\sqrt{4 \al_{n}^2 \beta^2 d (a +z)+\left(\al_{n}^2 d (1-\beta )+\beta (a +z)\right)^2}+\al_{n}^2 (\beta-1) d+\al_{n} \beta-\beta (a +z)}{2 \beta (a +z)}.\label{delta}
 \end{align}
 \line(1,0){470}
\end{figure*}
Note that  the general expression of $t^{'}_{i}$ for $i=1,2,3$ is given by 
 \begin{align}
t^{'}_{i}=\frac{\beta d^2 \delta^{2}(\delta+1)^2 k_i}{a^4 \left(\beta (\delta+1)^2-d^4 \delta^2\right)}, \nn
\end{align}
since  $e^{'}_{k_{i}}=e^{'}_{\bK}=\beta d \delta^2 k_i/(\al_{n}^2 (\beta - (d^2 \delta^2)/(1 + \delta)^2)) $. Here, $k_{1}=(1  - \al_{n}^{2}d)$, $k_{2}=\al_{n}^{2} d$, and $k_{3}=1$. 
\endproof

%
%

\section{Numerical Results}\label{results}
The purpose of this section is to present some representative numerical examples enabling the study of the time variation of the channel due to the effects of phase noise and  user mobility on the performance of massive MIMO systems with MRT and RZF precoders. The metrics under study are the achievable sum-rates and the required transmit power achieving a certain user rate. Note that the achievable sum-rates are described by means of the SINRs, given by Theorems~\ref{theorem:DLagedCSIMRT} and~\ref{theorem:DLagedCSIRZF}. In addition, the correctness of the proposed results is validated by Monte-Carlo simulations. Notably, despite that the analytical results are obtained by  assuming that $M,K \to \infty$, they coincide with the simulations even for finite values of $M$ and $K$. Actually, this is a known observation in the literature concerning the DEs~\cite{Couillet2011,Hoydis2013,Truong2013,Papazafeiropoulos2014,Papazafeiropoulos2014WCNC}.
\subsection{Simulation Setup}
Given that the interest of this work is based  on the investigation of the impact of practical channel impairments, the setting of our scenario includes long-term evolution (LTE) system specifications~\cite{Marzetta2010}. Specifically,  the simulation setup considers a single cell with a radius of $R = 1000$ meters, and  a guard range of $r_{0} = 100$ meters specifying the distance between the nearest user and the BS. The BS, comprised of  $M$ antennas, broadcasts to $K$ users that are  uniformly distributed within the cell. According to the system model, the channel vector between the BS and the UE $k$ in the $n$th time slot describes the large-scale fading and path loss or spatial correlation by means of $\bR_{k}$. Herein,  $\bR_{k}$ describes  large-scale fading  modelled as $\bR_{k}=l_{k}\Id_{M}$ with $l_{k}=q_{k}/\left( r_{k}/r_{0} \right)^{\upsilon}$. Especially, $q_k$ is a log-normal random variable with standard deviation $\sigma$ ($\sigma = 8$ dB) expressing the shadow-fading effect, $r_{k}$ denotes the distance between UE $k$ and the BS, and $\upsilon$ ($\upsilon = 3.8$) is the path loss exponent. The uplink and downlink powers are $p_{\mathrm{u}}=p_{\mathrm{d}}=46$dBm, while the thermal noise density is $-174$ dBm/Hz. The length of the training duration is $\tau=K$ symbols, and the phase noises  at the  BS and user LOs are simulated as discrete Wiener processes given by~\eqref{phaseNoiseBS} and~\eqref{phaseNoiseuser}, with increment variances in the interval $\operatorname{0\deg-2\deg}$~\cite{Colavolpe2005}. The coherence time is $T_{\mathrm{c}}=1/4 f_{\mathrm{D}}=1~\mathrm{ms}$, where $f_{\mathrm{D}}=250~\mathrm{Hz}$ is the Doppler spread corresponding to a relative velocity of  $135$ km/h between the BS and the users, if  the  center frequency is assumed to be $f_\mathrm{c}= 2~\mathrm{GHz}$. Moreover, given that the bandwidth for LTE-A is $\mathrm{W}=20\mathrm{MHz}$, the symbol time is $T_s=1/(2\mathrm{W})=0.025~\mathrm{\mu s}$. In order to account for fast varying channels, where high mobility occurs, the coherence block is assumed to include $T=196$ channel uses corresponding to the coherence bandwidth ${B}_{\mathrm{c}}=196~\mathrm{KHz}$.

One useful metric, providing the means for study the considered system, is
\begin{align} \label{eq num 1}
 \mathcal{S} \triangleq\sum_{k=1}^K
 \bar{R}_{k},
\end{align}
where $\mathcal{S}$ is the sum-rate and $\bar{R}_{k}$ is the DE rate of UE $k$ given by~\eqref{DeterministicSumrate}, and 
Theorems~\eqref{theorem:DLagedCSIMRT} and~\eqref{theorem:DLagedCSIRZF} for the cases of MRT and RZF, respectively.

In the following figures, the red and black line patterns correspond to the rates with RZF and MRT precoding for various phase noise nominal values, respectively. Thus, the ``solid'', ``dash'', and ``dot'' lines designate the analytical results with no channel aging, $\phi_{m,n}=0\deg,
 \varphi_{k,n}=2\deg$, and $\phi_{m,n}= \varphi_{k,n}=2\deg$, respectively.   The bullets represent the simulation results. Finally,  the green ``solid'' and ``dot'' lines, where applicable, depict a scenario with channel aging but not phase noise in the cases of MRT and RZF, respectively.
\subsection{Rate Comparison between MRT and RZF}
\begin{figure}[t]
    \centering
    \centerline{\includegraphics[width=0.5\textwidth]{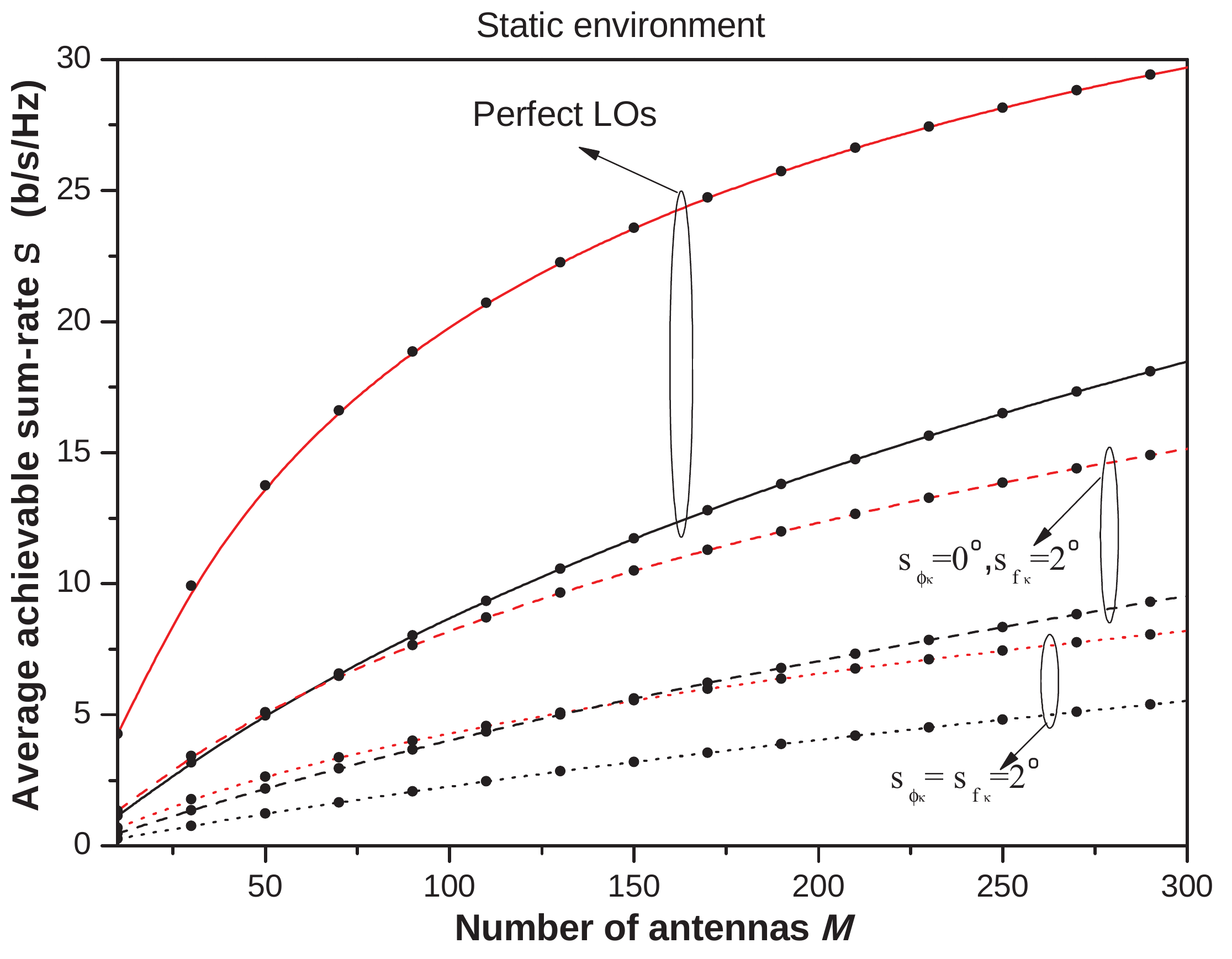}}
    \caption{Simulated and DE downlink sum-rates with MRT and RZF precoders in a static environment versus the number of BS antennas for  various values of phase noise. Red and black lines correspond to the theoretical sum-rates with RZF and MRT precoding, respectively, while the  black bullets refer to the simulation results.}
    \label{fig:1}
\end{figure}

The rate comparison between MRT and RZF considers the impact of both Doppler shift and phase noise. Clearly, the RZF outperforms the MRT for the scenarios considered. Especially, Fig.~\ref{fig:1} depicts the achievable rates by varying the number of BS antennas for different values of phase noise, when the users are assumed to be static ($v=0~\mathrm{Km/h}$), or equivalently, in the case of no Doppler shift ($f_{\mathrm{D}}=0~\mathrm{Hz}$). By increasing the hardware imperfection in terms of phase noise, the sum-rate decreases. As far as the RZF is concerned, when the phase noise at the user side has a variance of $\sigma_{\varphi_{k}}^{2}=2\deg$ the rate loss is $75\%$ with the BS having $M=30$ antennas, but when $M$ reaches $300$, the loss is smaller, i.e.,  $52\%$.  
Moreover, when the phase noise increases, the sum-rate saturates faster. Similar conclusions can be made in the case of MRT precoding. The perfect match between Monte Carlo simulations and analytical results validates the analytical results.

On the other hand, Fig.~\ref{fig:2} provides the variation of $\mathcal{S}$ versus the normalized Doppler shift $f_{\mathrm{D}}T_{\mathrm{s}}$ when $M=60$ for different values of phase noise starting from perfect LOs (no phase noise) to high phase noise ($\sigma_{\phi_{k}}^{2}=\sigma_{\varphi_{k}}^{2}=2\deg$). It is revealed that the effect of the Doppler shift on the achievable rate is more detrimental than that of phase noise. In fact, for low velocities in the order of $30~\mathrm{km/h}$ equivalent to $f_{\mathrm{D}}T_{\mathrm{s}}\approx 0.2$, the degradation due to phase noise starts to become  insignificant, but then the achievable rate can become so low that it is inadequate for practical applications, i.e., it is not worthy to investigate the impact of phase noise in high mobility environments. Hence, the higher the mobility, the less important role the phase noise plays. In the same figure, the straight lines illustrate the sum-rate with imperfect CSI, but with no user mobility and with perfect LOs at both BS and user ends.  Furthermore, it is evident that as phase noise increases, the performance worsens. However, the loss due to phase noise is prominent in static environments. 
\begin{figure}[t]
    \centering
    \centerline{\includegraphics[width=0.5\textwidth]{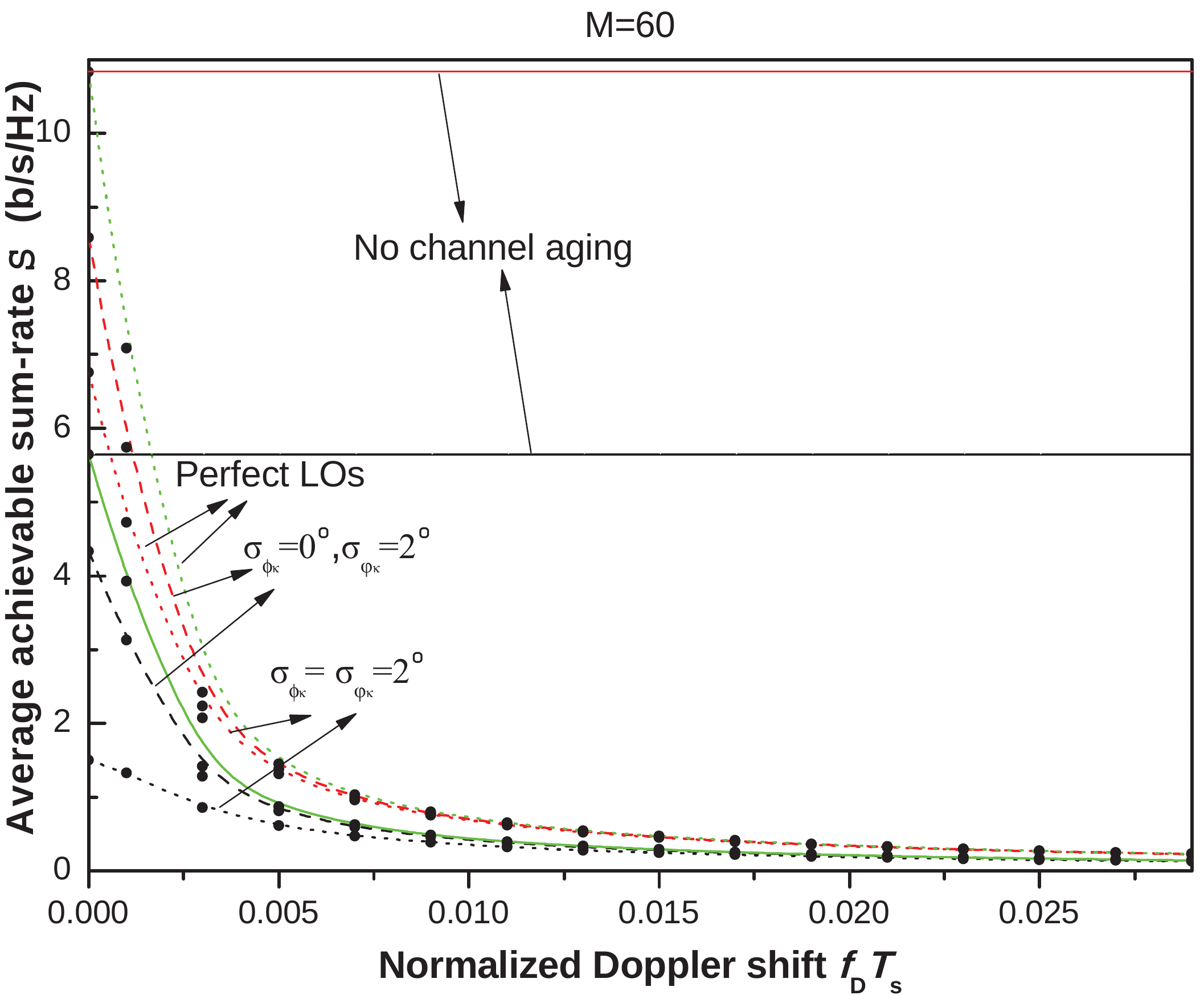}}
    \caption{Simulated and DE downlink sum-rates with MRT and RZF precoders when $M=60$ as a function of  the normalized Doppler shift for  various values of phase noise. Red and black lines correspond to the theoretical sum-rates with RZF and MRT precoding, respectively, while the black bullets refer to the  simulation results. The green ``solid'' and ``dot'' lines mirror a scenario with channel aging but not phase noise in the cases of MRT and RZF, respectively. Lines parallel to x-axis represent scenarios with no channel aging.}
    \label{fig:2}
\end{figure}
\subsection{Required Transmit Power Comparison between MRT and RZF}
Regarding the required transmit power achieving a specific rate per user equal to $1$ bit/s/Hz, it is quite insightful to investigate how it changes by varying the number of BS antennas, the amount of Doppler shift, and  the severity of phase noise.

\begin{figure}[t]
    \centering
    \centerline{\includegraphics[width=0.5\textwidth]{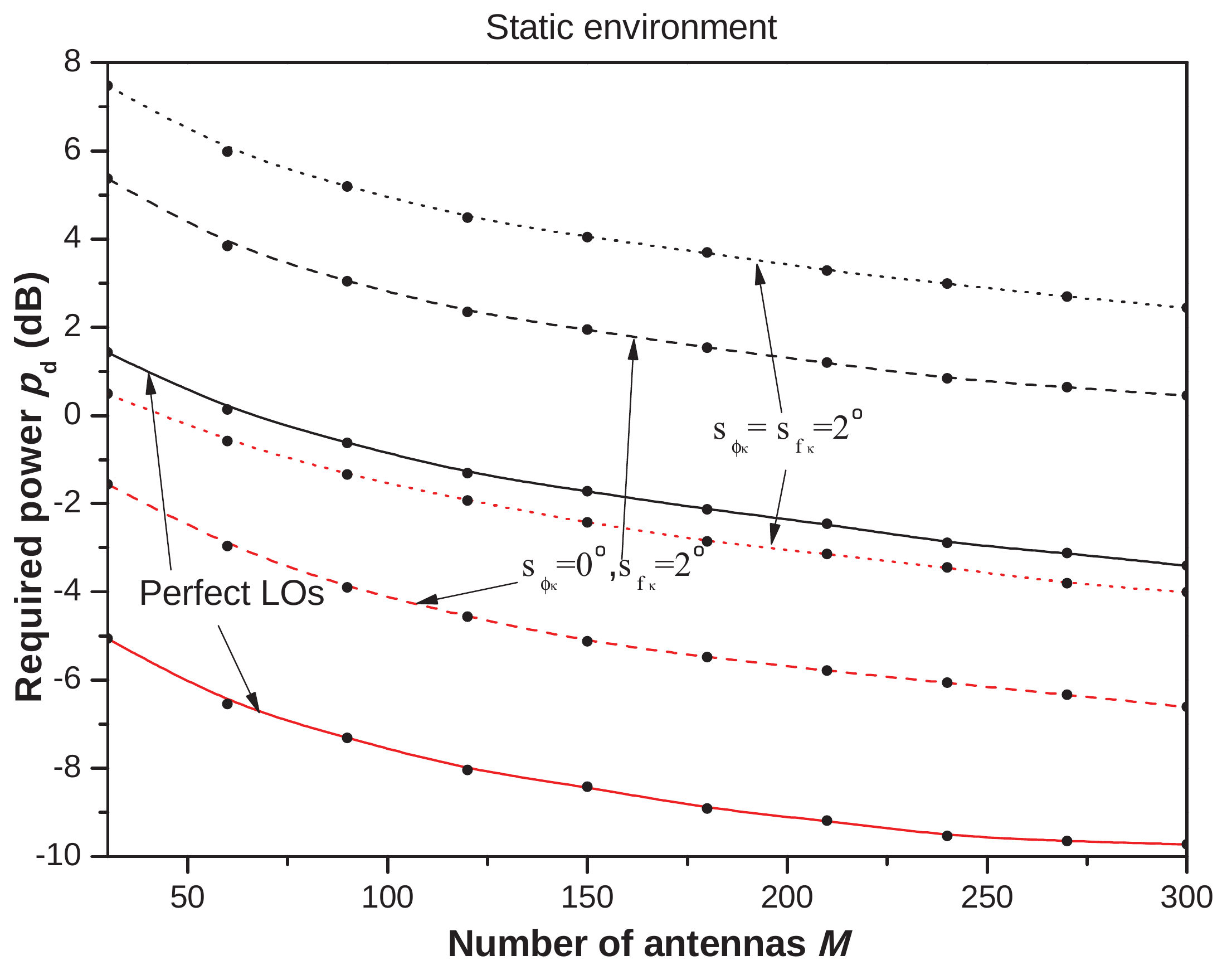}}
    \caption{Required transmit power  to achieve $1$ bit/s/Hz per user  with MRT and RZF precoders in a static environment versus the number of BS antennas and various values of phase noise. Red and black lines correspond to the theoretical sum-rates with RZF and MRT precoding, respectively, while the black bullets refer to the simulation results.}
    \label{fig:3}
\end{figure}

Hence, Fig.~\ref{fig:3} illustrates the variation of the transmit power $p_{\mathrm{d}}$ versus the number of BS antennas in a static environment ($f_{\mathrm{D}}T_{\mathrm{s}}=0$). Specifically,  $p_{\mathrm{d}}$ decreases considerably when we increase $M$. Especially, a closer observation shows a reduction in the transmit power by approximately 1.5dB  after doubling the number of BS antennas, which agrees with previously known results e.g.,~\cite{Ngo2013}. Notably, the more severe is the phase noise at the user, the more the required transmit power should be. 

\begin{figure}[t]
    \centering
    \centerline{\includegraphics[width=0.5\textwidth]{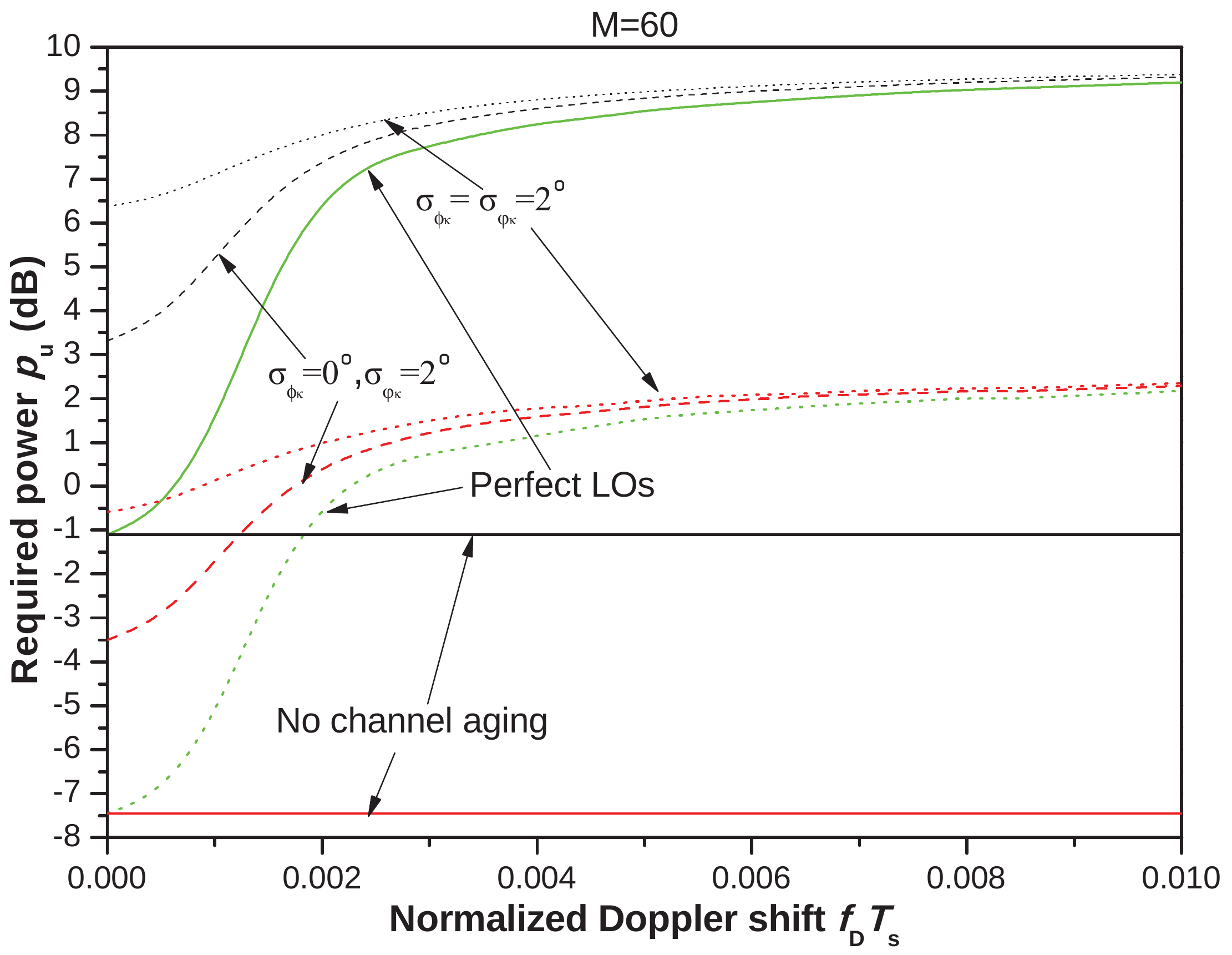}}
    \caption{Required transmit power  to achieve $1$ bit/s/Hz per user with MRT and RZF precoders  when $M=60$ as a function of the normalized Doppler shift for  various values of phase noise.  Red and black lines correspond to the theoretical sum-rates with RZF and MRT precoding, respectively. The green ``solid'' and ``dot'' lines mirror a scenario with channel aging but not phase noise in the cases of MRT and RZF, respectively. Lines parallel to x-axis represent scenarios with no channel aging.}
    \label{fig:4}
\end{figure}

Fig.~\ref{fig:4} depicts $p_{\mathrm{d}}$ versus the varying normalized Doppler shift $f_{\mathrm{D}}T_{\mathrm{s}}$ and phase noise,  when $M=60$. In particular, the straight lines represent the required transmit power with no channel aging. As a result, these do not depend on $f_{\mathrm{D}}T_{\mathrm{s}}$ or phase noise. However, as the Doppler shift increases,  $p_{\mathrm{d}}$ becomes higher, and very soon (low velocities), it saturates to a constant. In other words, after a specific value of  $f_{\mathrm{D}}T_{\mathrm{s}}$, any increase of  $p_{\mathrm{d}}$ cannot achieve any benefit. Furthermore, the higher the phase noise, the faster the saturation ensues.

\subsection{Extension to Multi-Cell Large MIMO Systems}
Herein, we focus on the impact of channel aging in the practical case of a hexagonal cellular system. Especially, we consider the downlink of a cellular MIMO system with $L$ cells
operating under the same frequency band, i.e., the system is impaired by pilot contamination. Actully, each cell can be assumed as an instance of the single-cell setting provided in Section II. The received signal at the $k$th user, located at the $i$th cell, is given by
\begin{align}
\!\!y_{ik,n}=\sqrt{p_{\mathrm{d}}}\sum_{l=1}^{L}\bh^\H_{lik,n}\bTheta_{lk,n}\bs_{l,n}+z_{ik,n}\label{eq:DLreceivedSignal}
\end{align}
where  $\bs_{l,n}$ denotes the unit-norm signal vector transmitted by the BS. Moreover, we have assumed that the BS transmits data to all the users with the same power $p_{\mathrm{d}}$, and $z_{ik,n} \sim \cC\cN(0,\sigma_{k}^{2})$ is 
complex Gaussian noise at user $k$, found in the $i$th cell.

The LMMSE estimator of $\bg_{ik,0}$, obtained during the training phase, is
 \begin{align}\label{estimatedChannel}
 \hat{\bg}_{ik,0}=\left(  \Id_{M}+\frac{\sigma_{\mathrm{b}}^{2}}{p_{\rp}}\sum_{l}\bR_{lik}^{-1}\right)^{-1}\tilde{\by}_{iik,0}^{\rp},
\end{align}
where $\sigma_{\mathrm{b}}^{2}$ is the variance of the post-processed noise at
base station, $\tilde{\by}_{ik,0}^{\rp}$ is a noisy observation of the effective channel from user $k$ to the $i$th BS, and $p_{\mathrm{p}}=\tau p_{\mathrm{u}}$  with $p_{\mathrm{u}}$   being the power per user in the uplink data transmission phase.

Given that the focal point of this work is to conduct an investigation of a generalized channel aging model, and  shed deeper light on channel aging as well as reveal new properties, we present only simulation results and not the corresponding deterministic equivalent analysis, which is straightforward and will distract the reader from the main objective of this part. In particular, we consider the same design parameters with the setup of the single cell, but now we employ $L=7$ cells. Moreover, the average sum-rate is going to be investigated for the central cell, while it has to be mentioned that in the multi-cell setting the large-scale fading depends on the distance of not only the distance of $k$th user from its associated BS, but also from the neighbor BSs. In such scenario, the equivalent SISO model for UE $k$ in the central cell has 
desired signal power and interference plus noise power given by 
 \begin{align}
S_{ik,n}
=  \lambda\Big|\EE\left[\bg^\H_{iik,n}\widetilde{\bTheta}_{ik,n}{\bff}_{ik,n}\right]  \Big|^{2}\label{eq:DLgenSignalPower5}
\end{align}
and 
\begin{align}
I_{ik,n}
&=   \lambda\var\left[\bg^\H_{iik,n}\widetilde{\bTheta}_{ik,n} \bff_{ik,n}\right]  +   \frac{\sigma_{k}^{2}}{p_{\mathrm{d}}} \nn\\
&+    \sum_{(l,m \ne i,k)}     \lambda\EE\bigg[\Big|\bg^\H_{lim,n}\widetilde{\bTheta}_{lm,n} \bff_{lm,n}\Big|^{2}\bigg],   \label{eq:DLgenIntfPower5}
\end{align}
where $\widetilde{\bTheta}_{ik,n}=\Delta{\bTheta}_{ik,n}$ and ${\bff}_{ik,n}$ represent the accumulated phase noise and the precoder applied by the $i$th BS.
\begin{figure}[t]
    \centering
    \centerline{\includegraphics[width=0.5\textwidth]{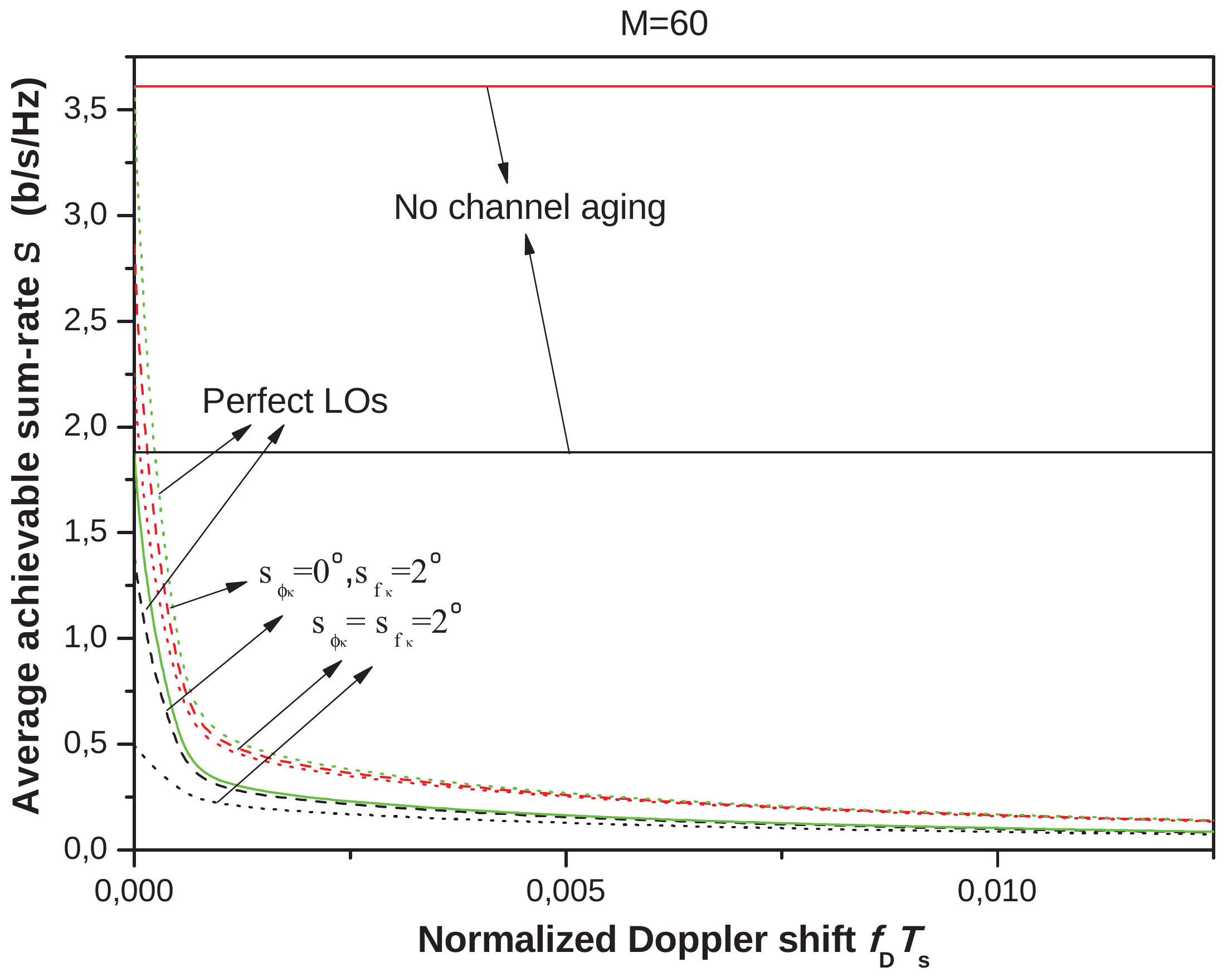}}
    \caption{Simulated downlink sum-rates with MRT and RZF precoders in a cellular setting with $L=7$ when $M=60$ as a function of  the normalized Doppler shift for  various values of phase noise. Red and black lines correspond to the simulated sum-rates with RZF and MRT precoding, respectively. The green ``solid'' and ``dot'' lines mirror a scenario with channel aging but not phase noise in the cases of MRT and RZF, respectively. Lines parallel to x-axis represent scenarios with no channel aging.}
    \label{fig:2}
\end{figure}

Obviously, the difference between the single-cell and multi-cell analyses is limited to the term describing the inter-cell interference, or in other words, the pilot contamination. Fig.~5 illustrates the sum-rate of the $i$th (central) cell versus the normalized Doppler shift. As can be seen, the impact of channel aging on the inter-cell interference is similar to the intra-cell interference, and results only in an extra degradation of the  system performance. In addition, the system is more sensitive to the normalized Doppler shift. In other words, the more severe the channel aging, and basically the user mobility, the higher the degradation of the achievable  sum-rate because it affects the inter-cell interference term.

\section{Conclusions}
In this work, we  modeled channel aging by incorporating both effects of Doppler shift coming from user mobility and phase noise due to circuitry imperfections of LOs, since both the effects contribute to the time variation of the effective channel multiplying/affecting the transmitted data. Given this novel integrated framework, we provided a joint channel-phase noise estimate. Next, the new CSI model was exploited to construct the MRT  and the RZF precoders, and derived the DEs of the corresponding downlink SINRs. As attested by Monte Carlo simulations, these DEs are tight approximations for the rate performance of the studied system. In addition, the numerical results showed that the degradation due to Doppler shift dominates against phase noise. As a result, the detrimental effect of phase noise is meaningful only in low mobility conditions. 
Notably, the use of massive MIMO systems should be preferred even in general channel aging conditions,  where, in addition, RZF behaves better than MRT as expected. Finally, we showed that in the case of MRT precoding, the required transmit power per user to achieve a certain rate can be scaled down at most by the inverse of the square root of the number of antennas, while a similar result should hold for RZF precoding as well.

\begin{appendices}
\section{Useful Lemmas}
\begin{lemma}[Matrix inversion lemma (I) {\cite[Eq.~2.2]{Bai1}}]\label{lemma:inversion}
\\Let $\bB\in\CC^{M\times M}$ be Hermitian  invertible. Then, for any vector $\bx\in\CC^{M}$, and any scalar $\tau\in\CC^{M}$ such that $\bB+\tau\bx\bx^\H$ is invertible,
\begin{align}
\bx^\H(\bB+\tau\bx\bx^\H)^{-1}=\frac{\bx^\H\bB^{-1}}{1+\tau\bx^\H\bB^{-1}\bx}.\nn
\end{align}
 \end{lemma}

\begin{lemma}[Matrix inversion lemma (II) {\cite[Lemma~2]{Hoydis2013}}]\label{lemma:inversion2}
\\Let $\bB\in\CC^{M\times M}$ be Hermitian  invertible. Then, for any vector $\bx\in\CC^{M}$, and any scalar $\tau\in\CC^{M}$ such that $\bB+\tau\bx\bx^\H$ is invertible,
\begin{align}
(\bB+\tau\bx\bx^\H)^{-1}=\bB-\frac{\bB^{-1}\tau\bx\bx^\H\bB^{-1}}{1+\tau\bx^\H\bB^{-1}\bx}.\nn
\end{align}
 \end{lemma}

\begin{lemma}[Rank-1 perturbation lemma {\cite[Lemma~2.1]{Bai2}}]
\\Let $z\in<0$, $\bB\in\CC^{M\times M}$, $\bB\in\CC^{M\times M}$ with $\bB$ Hermitian  nonnegative-definite, and $\bx\in\CC^{M}$. Then,
\begin{align}
|\tr\left((\bB-z\Id_M)^{-1} -(\bB+\bx\bx^\H-z\Id_M)^{-1}\bB\right)|\leq\frac{\|\bB\|}{|z|}.\nn
\end{align}
 \end{lemma}

\begin{lemma}[{\cite[Thm. 3.7]{Hoydis2013}},{\cite[Lem. 1]{Truong2013}}]\label{lemma:asymptoticLimits}
Let $\bB \in \bbC^{M \times M}$ with uniformly bounded spectral norm (with respect to $M$). Consider $\bx$ and $\by$, where $\bx, \by \in \bbC^{M}$, $\bx \sim \cC\cN(\b0, \bPhi_{x})$ and $\by  \sim \cC\cN(\b0, \bPhi_{y})$, are mutually independent and independent of $\bB$. Then, we have
\begin{align}
\frac{1}{M}\bx^{\H}\bB\bx - \frac{1}{M}\tr \bB\bPhi_{x} & \xrightarrow[ M \rightarrow \infty]{\mbox{a.s.}} 0 \label{eq:oneVector}\\
\frac{1}{M}\bx^{\H}\bB\by & \xrightarrow[ M \rightarrow \infty]{\mbox{a.s.}} 0 \label{eq:twoVector}\\
\EE\!\!\left[\left|\left(\frac{1}{M}\bx^{\H}\bB\bx\right)^{2}\!\! - \!\left(\frac{1}{M}\tr \bB \bPhi_{x} \right)^{2} \right|\right] \!\!&\xrightarrow[ M \rightarrow \infty]{\mbox{a.s.}}  0\label{eq:squared}\\
\frac{1}{M^{2}} |\bx^{\H}\bB\by|^{2} - \frac{1}{M^{2}} \tr \bB \bPhi_{x} \bB^{\H} \bPhi_{y}  & \xrightarrow[ M\rightarrow \infty]{\mbox{a.s.}} 0. \label{eq:twoVectorGeneral}
\end{align}
\end{lemma}

\begin{lemma}[{\cite[p. 207]{Tao2012}}]\label{lemma:asymptoticproduct}
Let $\bA\mathrm{,}~\bB\in \mathcal{C}^{M\times M}$ be freely independent random
matrices  with uniformly bounded spectral norm for all $M$. Further, let all the moments of the entries of $\bA\mathrm{,}~\bB$ be finite. Then,
\begin{align}
\frac{1}{M}\tr \bA\bB-\frac{1}{M}\tr \bA \frac{1}{M}\tr \bB \xrightarrow[ M\rightarrow \infty]{\mbox{a.s.}} 0.
\end{align}
\end{lemma}

\begin{Theorem}[{\cite[Theorem 1]{Wagner2012}}]\label{th:detequ}
Let $\bL\in\CC^{M\times M}$ and $\bS\in\CC^{M\times M}$ be Hermitian  nonnegative definite matrices, and let $\bH\in\CC^{M\times K}$ be a random matrix with columns $\bv_k\sim \cC \cN\left(0,\frac{1}{M}\bR_k\right)$. Assume that $\bL$ and the matrices $\bR_k$, $k=1,\dots,K$, have uniformly bounded spectral norms (with respect to $M$). Then, for any $\rho>0$,
\begin{align*}
\frac1M\tr\bL\left(\bH\bH^\H +\bS +\rho\Id_M\right)^{-1} - \frac1M\tr\bL\bT(\rho) \xrightarrow[ M_{\rt} \rightarrow \infty]{\mbox{a.s.}} 0,
\end{align*}
where $\bT(\rho)\in\CC^{M\times M}$ is defined as
$$ \bT(\rho) = \left(\frac1M\sum_{k=1}^K\frac{\bR_k}{1+e_k(\rho)}  +\bS + \rho\Id_M\right)^{-1},$$
and the elements of $\boldsymbol{e}(\rho)=\left[e_1(\rho)\cdots e_K(\rho)\right]^\T$ are defined as $ e_k(\rho)  = \lim_{t \to \infty}e_k^{(t)}(\rho),$ where for $t=1,2,\ldots$
\begin{align}
\!\!\!e_k^{(t)}(\rho)\!=\!\frac1M\tr\bR_k\!\left(\!\!\frac1M\!\!\sum_{j=1}^{K}\!\!\frac{\bR_j}{1+e_j^{(t-1)}(\rho)}+\bS+\rho\Id_M\!\!\right)\!\!^{-1}
\end{align}
with initial values $e_k^{(0)}(\rho)=\frac{1}{\rho}$ for all $k$.
\end{Theorem}
\vspace{5pt}

\begin{Theorem}[{\cite[Theorem 2]{Hoydis2013}}]\label{th:detequder}
Let $\bTheta\in\CC^{M\times M}$ be a Hermitian  nonnegative definite matrix with uniformly bounded spectral norm (with respect to $M$). Under the same conditions as in Theorem~\ref{th:detequ}, we have  
\begin{align*}
\frac1M\tr\bL\left(\bH\bH^\H + \bS+\rho\Id_M\right)^{-1}&\bK\left(\bH\bH^\H  +\bS+\rho\Id_M\right)^{-1}\nn\\
&- \frac1M\tr\bL\bT^{'}(\rho) \xrightarrow[]{\text{a.s.}} 0,
\end{align*}
where $\bT^{'}(\rho)\in\CC^{M\times M}$ is defined as
$$ \bT^{'}(\rho) = \bT(\rho)\bK \bT(\rho) + \bT(\rho)\frac1M\sum_{k=1}^K\frac{\bR_k e^{'}_k(\rho) }{\left(1+e_k(\rho)\right)^2}\bT(\rho)$$
with $\bT(\rho)$ and $\boldsymbol{e}_k(\rho)$ as defined in Theorem~\ref{th:detequ} and $\boldsymbol{e}^{'}(\rho) = \left[e^{'}_1(\rho)\cdots e^{'}_K(\rho)\right]^\T$ given by
\begin{align}
 \boldsymbol{e}^{'}(\rho) &= \left(\Id_K - \bJ(\rho)\right)^{-1}\bv(\rho).
\end{align}
The elements of $\bJ(\rho)\in\CC^{K\times K}$ and $\bv(\rho)\in\CC^{K}$ are defined as
\begin{align}
[\bJ(\rho)]_{kl} = \frac{\frac1M\tr\bR_k\bT(\rho)\bR_l\Tm(\rho)}{M\left(1+e_k(\rho)\right)^2}\nn
\end{align}
and
\begin{align}
[\bv(\rho)]_k = \frac1M\tr\bR_k\Tm(\rho)\bK\Tm(\rho).\nn
\end{align}

\end{Theorem}

\section{Proof of Proposition~\ref{LMMSE}}\label{proposition1} 
During the training phase,  users transmit mutually orthogonal training sequences consisting of $\tau$ symbols, and it is assumed that the channel remains  constant during this phase.  In particular,  the pilot sequences can be represented by  $\bPsi = [\bpsi_{1}; \cdots; \bpsi_{K}] \in \bbC^{K \times \tau}$ with $\bPsi$ normalized, i.e., $\bPsi\bPsi^\H = \bI_K$. Given that the channel estimation takes place at time $0$, the received signal at the BS is written as
\begin{align}\label{eq:Ypt}
\bY_{\rp,0}= & \sqrt{p_{\rp}}\bTheta_{k,0} \bH_{0}\bPsi + \bZ_{\rp,0},
\end{align}
where $p_{\rp}$ is the common average transmit power for all users, $\bTheta_{k,0}=\mathrm{diag}\left\{ e^{j \theta_{k,0}^{(1)}}, \ldots, e^{j \theta_{k,0}^{(M)}} \right\}$ is the phase noise because of the BS and user $k$ LOs at time $0$, and $\bZ_{\rp,0} \in \bbC^{M \times \tau}$ is the spatially white additive Gaussian noise matrix at the  BS   during the training phase. Note that $ \theta_{k,0}^{(m)}=\phi_{m,0}+\varphi_{k,0},~m=1,\ldots, M$. By correlating the received signal with the training sequence $\frac{1}{\sqrt{p_{\rp}}}\bpsi^\H_k$ of user $k$, and by substituting $\bg_{k,0}=\bTheta_{k,0} {\bh_{k,0}}$ the BS obtains
\begin{align}
\tilde{\by}_{k,0}^{\rp}
= \bg_{k,0}  + {\frac{1}{\sqrt{p_{\rp}}}\tilde{\bz}_{\rp,0}},\label{eq:Ypt3}
\end{align}
where $\tilde{\bz}_{\rp,0}\triangleq \bZ_{\rp,0}\bpsi^\H_{k}\sim \cC\cN(\b0,\sigma_{\mathrm{b}}^{2}\bI_{M})$.  After applying the MMSE estimation method~\cite{Kay}, the  effective channel estimated at the BS  can be written as~\eqref{estimatedChannel}.
Employing the orthogonality principle, the  channel decomposes as
\begin{align}
\bg_{k,0} = \hat{\bg}_{k,0} + \tilde{\bg}_{k,0},\label{eq:MMSEorthogonality}
\end{align}
where $\hat{\bg}_{k,0}$ is distributed as $\cC\cN\left(\b0,\bD_{k}\right)$ with $\bD_{k}=\left(  \Id_{M}+\frac{\sigma_{b}^{2}}{p_{\rp}}\bR_{k}^{-1}\right)^{-1}\bR_{k}$, and $\tilde{\bg}_{k,0} \sim \cC\cN(\b0, \bR_{k} - \bD_{k})$ is the channel estimation error vector. Note that $\hat{\bg}_{k}$ and $\tilde{\bg}_{k}$ are statistically independent because they are uncorrelated and jointly Gaussian.
 \section{Proof of Theorem~\ref{theorem:DLagedCSIMRT}}\label{theorem2}
 We consider the desired signal power 
\begin{align}
S_{k,n}
&= \frac{\lambda}{M}  \Big|\EE\left[{\bg}^\H_{k,n}\widetilde{\bTheta}_{k,n}   \bA_{n} \hat{\bg}_{k,0} \right]\!\!\Big|^{2}.\label{normalizedMRT}
\end{align}
First, we  derive the DE of the normalization parameter $\lambda$. Hence, the normalization parameter can be written by means of~\eqref{eq:lamda} and~\eqref{eq:precoderMRT} as
\begin{align}
\lambda&=
\frac{K}{\EE\Big[\frac{1}{M}\tr \bA_{n}{\hat{\bG}}_{0}{\hat{\bG}}_{0}^{\H} \bA_{n}   \Big]} \nn\\
&= \frac{K}{\EE\Big[\sum_{i=1}^{K}\frac{1}{M}\bA_{n}^{2}  \hat{\bg}_{k,0} \hat{\bg}_{k,0}^{\H}  \Big]}\nn\\
&\asymp \left( \frac1K\sum_{i=1}^{K}\frac{1}{M}\tr \bA_{n}^{2}\bD_{k} \right)^{-1},\label{desired1MRT}
\end{align}
where we have applied Lemma~\ref{lemma:asymptoticLimits}.
As far as the other term in~\eqref{normalizedMRT} is concerned, we obtain
\begin{align}
 \frac{1}{M}\EE\!\left[{\bg}^\H_{k,n}\widetilde{\bTheta}_{k,n}  \bA_{n} \hat{\bg}_{k,0} \right]
 &\!=\! \frac{1}{M}\EE\!\left[(\hat{\bg}^\H_{k,0}\bA_{n}  \!+ \!\tilde{\bee}^\H_{k,n} ) \widetilde{\bTheta}_{k,n}\bA_{n} \hat{\bg}_{k,0} \right]\nn\\
 &=\frac{1}{M}\EE\Big[{\hat{\bg}^\H_{k,0}\bA_{n}\widetilde{\bTheta}_{k,n} \bA_{n}\hat{\bg}_{k,0} }\Big],\nn\\
 &\asymp\frac{1}{M}\EE\left[\tr\bA_{n}\widetilde{\bTheta}_{k,n}\bA_{n}\bD_{k}\label{desired4MRT}\right]\\
 &\asymp \frac{e^{-\left( \sigma_{\varphi_{k}}^{2}+\sigma_{\phi}^{2} \right)n}}{M}\tr\bA_{n}^{2}\bD_{k},\label{desired2MRT}
\end{align}
where in~\eqref{desired4MRT} and~\eqref{desired2MRT}, we have applied Lemmas~\ref{lemma:asymptoticLimits} and~\ref{lemma:asymptoticproduct}, respectively. Note that $\bA_{n}$ and $\widetilde{\bTheta}_{k,n}$ commute because both are diagonal matrices. In particular, regarding the CLO setting, it holds that $\frac{1}{M}\tr \widetilde{\bTheta}_{k,n}=e^{ 2 j \left( \varphi_{k, n}-\varphi_{k, 0} +\phi_{k, n}-\phi_{k, 0}\right)}$, while in the case of the SLOs setup, we obtain $\frac{1}{M}\tr \widetilde{\bTheta}_{k,n}=e^{2 j \left( \varphi_{k, n}-\varphi_{k, 0} \right)-\sigma_{\phi}^{2}n}$ or  $\frac{1}{M}\tr \widetilde{\bTheta}_{k,n}=e^{2 j \left( \varphi_{k, n}-\varphi_{k, 0} \right)}\frac{1}{M}\prod_{i=1}^{M}e^{-\sigma_{\phi_{i}}^{2}n}$, when the LOs obey to identical or non-identical statistics, respectively. Herein, we focus on the former scenario regarding the BS antennas. Especially, in such case their increment variance is $\sigma_{\phi}^{2}$. Application of the expectation operator to  $\frac{1}{M}\tr \widetilde{\bTheta}_{k,n}$ gives~$e^{-\left( \sigma_{\varphi_{k}}^{2}+\sigma_{\phi}^{2} \right)n}$ for both the  CLO and SLOs settings. Finally, after simple algebraic maipulations, we lead to ~\eqref{desired2MRT}. By denoting $\bar{S}_{k,n} = \lim_{M \rightarrow \infty} S_{k,n}$ the DE signal power, and by using~\eqref{desired1MRT} and~\eqref{desired2MRT}, we have
\begin{align}
\bar{S}_{k,n}\asymp  \bar{\lambda}e^{-2\left( \sigma_{\varphi_{k}}^{2}+\sigma_{\phi}^{2} \right)n}  {\delta}_{k}^{2},\label{eq:theorem4.5MRT}
\end{align}
where ${\delta}_{k}=\frac{1}{M}\tr\bA_{n}^{2}\bD_{k}$.
Now, we proceed with the derivation of each term in~\eqref{eq:DLgenIntfPower}. The first term on the right hand side can be written as 
\begin{align}
\frac{1}{M} \var&\left[\bg^\H_{k,n} \widetilde{\bTheta}_{k,n} \bA_{n} \hatvg_{k,0} \right]\nn\\
&-\frac{1}{M^{2}} \EE\bigg[\Big|{\tilde{\bee}_{k,n}^{\H}\widetilde{\bTheta}_{k,n} \bA_{n}\hatvg
_{k,0} }\Big|^2\bigg]\xrightarrow[ M \rightarrow \infty]{\mbox{a.s.}} 0,
\end{align}
where the property of the variance operator $\mathrm{var}\left[ x \right]=\EE[x^{2}]- \EE^{2}[x] $ together with~\eqref{eq:GaussMarkov2} have been used. Lemmas~\ref{lemma:asymptoticLimits} and~\ref{lemma:asymptoticproduct} enable us to derive this DE as
\begin{align}
 \frac{\lambda}{M} \var\left[\bg^\H_{k,n} \widetilde{\bTheta}_{k,n}  \bA_{n} \hatvg_{k,0} \right] 
\asymp&  \frac{\bar{\lambda}}{M}{\delta}_{k}^{'}, \label{eq:theorem4.7MRT}
\end{align}
where ${\delta}_{k}^{'}=\frac{1}{M}\tr\bA_{n}^{2}\bD_{k} \left( \bR_{k} - \bA_{n}^{2}\bD_{k} \right)$.
Similarly, use of Lemmas~\ref{lemma:asymptoticLimits} and~\ref{lemma:asymptoticproduct} in the last term of~\eqref{eq:DLgenIntfPower} completes the proof. Thus, we have
\begin{align}
  \frac{\lambda}{M^{2}}\EE\Big[\left|\bg^\H_{k,n} \widetilde{\bTheta}_{k,n}\bA_{n} \hatvg_{i,0} \right|^{2}\Big] 
&\asymp \frac{\bar{\lambda}}{M^{2}} \tr\widetilde{\bTheta}_{k,n}\bA_{n}^{2}\bD_{i}\widetilde{\bTheta}_{k,n} ^{\H}\bR_{k},\nn\\
&=\frac{\bar{\lambda}}{M}\delta_{i}^{''},\label{lasttermMRT}
\end{align}
since $\bg_{k,n} $ and $\hatvg_{i,0} $ are mutually independent. Note here that $\delta_{i}^{''}=\frac{1}{M} \tr\bA_{n}^{2}\bD_{i} \bR_{k}$. 

 \section{Proof of Theorem~\ref{theorem:DLagedCSIRZF}}\label{theorem3}
First, we obtain  the DE of the normalization parameter $\lambda$. Taking into account for~\eqref{eq:lamda} and~\eqref{eq:precoderRZF}, and making a simple algebraic manipulation, we lead to
\begin{align}
\lambda&=
\frac{K}{\EE\Big[\tr{ \bSigma} \bA_{n}{\hat{\bG}}_{0}{\hat{\bG}}_{0}^{\H} \bA_{n} {\bSigma} \Big]} \label{eq:theorem4.6}\\
&= \frac{K}{\EE\Big[\tr {\bSigma}  - \tr \left( \Zm+ a M\Id_M\right)  {\bSigma}^{2} \Big]},
\end{align}
while its DE is 
\begin{align}
 \bar{\lambda}=\frac{K}{ \left(\frac{1}{M}\tr\Tm - \frac{1}{M}\tr \left(\frac{\Zm}{M} + a \Id_M\right) {\bC}\right)},\label{desired1}
\end{align}
where we have applied Theorems 1 and 2 for $\bK=\Id_M$.
Regarding the other term of the desired signal power, we have
\begin{align}
  &\EE\!\left[{\bg}^\H_{k,n} \widetilde{\bTheta}_{k,n}{\bSigma} \bA_{n} \hat{\bg}_{k,0} \right]
\!=\! \EE\!\left[(\hat{\bg}^\H_{k,0}\bA_{n}\!  + \!\tilde{\bee}^\H_{k,n} )\widetilde{\bTheta}_{k,n}{\bSigma} \bA_{n}\hat{\bg}_{k,0} \right]\label{desiredsig1}\nn\\
 &=\EE\Big[\frac{\hat{\bg}^\H_{k,0} \bA_{n} \widetilde{\bTheta}_{k,n}{\bSigma}_k\bA_{n} \hat{\bg}_{k,0} }{1+{\hatvg}^\H_{k,0}\bA_{n}  \widetilde{\bTheta}_{k,n}{\bSigma}_k \widetilde{\bTheta}_{k,n} ^{\H}\bA_{n}{\hatvg}_{k,0} }\Big],
\end{align}
where Lemmas~\ref{lemma:inversion} and~\ref{lemma:asymptoticLimits} have been applied in \eqref{desiredsig1}, while $ {\bSigma}_k $ is defined as
\begin{align}
 {\bSigma}_k\! &=\!\left(\widetilde{\bTheta}_{k,n} ^{\H}\bA_{n}   \hat{\bG}_{0} \hat{\bG}^\H_{0} \bA_{n}\widetilde{\bTheta}_{k,n} \!-\!\widetilde{\bTheta}_{k,n} ^{\H}\bA_{n}\hat{\bg}_{k,n}\hat{\bg}^\H_{k,n}\bA_{n}\widetilde{\bTheta}_{k,n}
 \right.\nn\\
 & \left.+ \bZ + M  a\Id_M\right)^{-1}.\nn
\end{align}
Exploiting Lemmas~\ref{lemma:asymptoticLimits},~\ref{lemma:asymptoticproduct}, and Theorem~\ref{th:detequ} gives
\begin{align}
   \EE\!\left[{\bg}^\H_{k,n}\widetilde{\bTheta}_{k,n}{\bSigma} \bA_{n} \hat{\bg}_{k,0} \right]\asymp \EE\!\left[\frac{\frac{1}{M}\tr\bA_{n}^{2}\bD_{k}\widetilde{\bTheta}_{k,n}\bT}{1+\frac{1}{M}\tr\bA_{n}^{2}\bD_{k}\bT}\right].\label{desired2}
\end{align}

Hence, use of~\eqref{desired1} and~\eqref{desired2} provides the DE signal power ${\bar{S}_{k,n}}$ as
\begin{align}
\bar{S}_{k,n}\asymp  \bar{\lambda}  \bigg(\frac{e^{-\left( \sigma_{\varphi_{k}}^{2}+\sigma_{\phi}^{2} \right)n}{\delta}_{k}}{1+{\delta}_{k}}\bigg)^{2},\label{eq:theorem4.5}
\end{align}
where ${\delta}_{k}=\frac{1}{M}\tr\bA_{n}^{2}\bD_{k}\bT$. Note that the manipulation regarding $\widetilde{\bTheta}_{k,n}$ follows a similar analysis to~\eqref{desired2MRT}.
We continue with the derivation of each term of~\eqref{eq:DLgenIntfPower}. Specifically, after applying  Lemmas~1 and~\ref{lemma:asymptoticLimits}, as well as~\eqref{eq:GaussMarkov2} to the first term in~\eqref{eq:DLgenIntfPower}, we have
\begin{align}
 &\var\left[\bg^\H_{k,n}\widetilde{\bTheta}_{k,n}  {\bSigma}\bA_{n} \hatvg_{k,0} \right]\nn\\
 & - \EE\bigg[\Big|\frac{\tilde{\bee}_{k,n}^{\H}   \widetilde{\bTheta}_{k,n} {\bSigma}_k \bA_{n}\hatvg
_{k,0} }{1+\hatvg^\H_{k,0}  \bA_{n} \widetilde{\bTheta}_{k,n} {\bSigma}_k \widetilde{\bTheta}_{k,n} ^{\H}\bA_{n} \hatvg_{k,0} }\Big|^2\bigg]\xrightarrow[ M \rightarrow \infty]{\mbox{a.s.}} 0,
\end{align}
which yields
\begin{align}
 \lambda \var\left[\bg^\H_{k,n} \widetilde{\bTheta}_{k,n} {\bSigma} \bA_{n} \hatvg_{k,0} \right] 
\asymp \bar{\lambda} \frac{\frac{1}{M}{\delta}_{k}^{'}}{\left(1+{\delta}_{k}\right)^2}, \label{eq:theorem4.7}
\end{align}
where we have used Theorems~\ref{th:detequ} and~\ref{th:detequder} as well as Lemmas~\ref{lemma:asymptoticLimits} and~\ref{lemma:asymptoticproduct}. Note that ${\delta}_{k}^{'}=\frac{1}{M}\tr\bA_{n}^{2}\bD_{k}{\Tm}^{'}$ and $\bK= \bR_{k} - \bA_{n}^{2}\bD_{k}$.
Next, we focus on the last term of~\eqref{eq:DLgenIntfPower}, where we make use of Theorems~\ref{th:detequ} and~\ref{th:detequder} as well as Lemmas~1 and~\ref{lemma:asymptoticLimits} as before. In particular, if $i \neq k$, we have
\begin{align}
& \EE\Big[\left|\bg^\H_{k,n}\widetilde{\bTheta}_{k,n}{\bSigma}\bA_{n} \hatvg_{i,0} \right|^{2}\Big] 
= \EE\bigg[\left|\frac{\bg^\H_{k,n}\widetilde{\bTheta}_{k,n} \bSigma_i \bA_{n} \hat{\bg}_{i,0}  }{1+\hatvg^\H_{i,0} \bA_{n} \bSigma_i \bA_{n} \hatvg_{i,0} }\right|^{2}\bigg]\nn\\
&= \EE\bigg[\frac{ \bg^\H_{k,n}\widetilde{\bTheta}_{k,n}\bSigma_i \bA_{n} \hatvg_{i,0} \hatvg_{i,0}^{\H}\bA_{n} \bSigma_i  \widetilde{\bTheta}_{k,n}^{\H}\bg_{k,n}}{\left( 1+\hatvg^\H_{i,0} \bA_{n} \bSigma_i\bA_{n} \hatvg_{i,0} \right)^{2} }\bigg]\nn\\
&\asymp \EE\bigg[\frac{ \bg^\H_{k,n}\widetilde{\bTheta}_{k,n}\bSigma_i \bA_{n}\bD_{i} \bA_{n} \bSigma_i\widetilde{\bTheta}_{k,n}^{\H}\bg_{k,n}}{\left( 1+\delta_{i}\right)^{2} }\bigg],\label{lastterm}
\end{align}
where we have taken into consideration that $\bg_{k,n} $ and $\hatvg_{i,0} $ are mutually independent. Unfortunately, upon inspecting~\eqref{lastterm}, we observe that $\bSigma_i$ is not independent of $\hat{\bg}_{k,0}$. For this reason, we use  Lemma~\ref{lemma:inversion2}, which gives
\begin{align}
\bSigma_i={\bSigma}_{ik}-\frac{{\bSigma}_{ik}\widetilde{\bTheta}_{k,n}^{\H}\bA_{n}\hatvg_{k,0}\hatvg_{k,0}^{\H}\bA_{n}\widetilde{\bTheta}_{k,n}{\bSigma}_{ik}}{1+\hatvg^\H_{k,0} \bA_{n}\widetilde{\bTheta}_{k,n} {\bSigma}_{ik}\widetilde{\bTheta}_{k,n}^{\H}\bA_{n} \hatvg_{k,0} },\label{eq:theorem2.I.51}
\end{align}which introduces a new matrix ${\bSigma}_{ik}$ to~\eqref{lastterm} defined as
\begin{align}
{\bSigma}_{ik}\!&=\!\left(\!\widetilde{\bTheta}_{k,n}^{\H}\bA_{n}\hat{\bG}_{0}\hat{\bG}_{0}^\H\bA_{n}\widetilde{\bTheta}_{k,n}\! -\!\widetilde{\bTheta}_{k,n}^{\H}\bA_{n}\hat{\bg}_{i,0}\hat{\bg}^\H_{i,0}\bA_{n}\widetilde{\bTheta}_{k,n} \right.\nn\\
&\left.\!-\widetilde{\bTheta}_{k,n}^{\H}\bA_{n}\hat{\bg}_{k,0}\hat{\bg}^\H_{k,0}\bA_{n}\widetilde{\bTheta}_{k,n}\!+\! \bZ\! + \!M a\Id_M\right)^{\!-\!1}.
\end{align}
By substituting~\eqref{eq:theorem2.I.51} into~\eqref{lastterm}, we obtain
\begin{align}
  \EE\Big[\left|\bg^\H_{k,n}\widetilde{\bTheta}_{k,n}{\bSigma}\bA_{n} \hatvg_{i,0} \right|^{2}\Big] 
&=\frac{{Q}_{ik}}{M\left(1+{\delta_{i}}\right)^{2}},\label{eq:theorem2.I.6}
\end{align}           
where ${Q}_{ik}$ is given in~\eqref{eq:theorem2.I.mu1} with $\widetilde{\bSigma}_{ik}=\widetilde{\bTheta}_{k,n}\bSigma_{ik}$.
\begin{figure*}
\begin{align}
{Q}_{ik}&= \bg^\H_{k,n}
\widetilde{\bSigma}_{ik} \bA_{n}^{2}\bD_{i}\widetilde{ \bSigma}_{ik}^{\H}  \bg_{k,n}\!+\!\frac{\left|  \bg^\H_{k,n}\widetilde{{\bSigma}}_{ik}\widetilde{\bTheta}_{k,n}^{\H}\bA_{n} \bg_{k,0}\right|^{2}\hat{\bg}^\H_{k,0} \bA_{n} \widetilde{{\bSigma}}_{ik} \bA_{n}^{2}\bD_{i}\widetilde{{\bSigma}}_{ik}^{\H}\bA_{n}\hat{\bg}_{k,0}}{\left( 1+\hatvg^\H_{k,0} \bA_{n}\widetilde{{\bSigma}}_{ik} \widetilde{\bTheta}_{k,n}^{\H}\bA_{n}\hatvg_{k,0} \right)^{2}}\nn\\
&-2\mathrm{Re}\left\{  \frac{\hatvg^\H_{k,0}\bA_{n}\widetilde{{\bSigma}}_{ik}\widetilde{\bTheta}_{k,n}^{\H} \bg_{k,n}\bg_{k,n}^{\H}\widetilde{{\bSigma}}_{ik}\widetilde{\bTheta}_{k,n}^{\H}\bA_{n}^{2}\bD_{i}\widetilde{{\bSigma}}_{ik}\widetilde{\bTheta}_{k,n}^{\H}\bA_{n}\hatvg_{k,0}}{1+\hatvg^\H_{k,0} \bA_{n}\widetilde{{\bSigma}}_{ik} \widetilde{\bTheta}_{k,n}^{\H}\bA_{n}\hatvg_{k,0}}\right\}.
 \label{eq:theorem2.I.mu1}
\end{align}
\line(1,0){470}
\end{figure*}
The DE of each term in~\eqref{eq:theorem2.I.mu1} is obtained as
\begin{align}
 \!\!\!\!\bg^\H_{k,n}\widetilde{\bTheta}_{k,n}\bSigma_{ik} \bA_{n}^{2}\bD_{i} \bSigma_{ik}  \widetilde{\bTheta}_{k,n} ^{\H}\bg_{k,n}&\asymp  \frac{1}{M^{2}}\tr \bR_{k}\bC^{''}
 \end{align}
\begin{align}
 \hat{\bg}^\H_{k,0}\bA_{n}\widetilde{\bTheta}_{k,n}{\bSigma}_{ik} \bA_{n}^{2}\bD_{i}{\bSigma}_{ik}\widetilde{\bTheta}_{k,n}^{\H}\bA_{n}\hat{\bg}_{k,0}
 &\asymp  \frac{1}{M^{2}}\tr \bA_{n}^{2}\bD_{k}\bC^{''}\nn\\
 &=\frac{ \delta_{k}^{''}}{M}
\end{align}
\begin{align}
\bg_{k,n}^{\H} \widetilde{\bTheta}_{k,n}{\bSigma}_{ik} \widetilde{\bTheta}_{k,n} ^{\H}&\bA_{n}^{2}\bD_{i} \widetilde{\bTheta}_{k,n}{\bSigma}_{ik} \widetilde{\bTheta}_{k,n}^{\H}\bA_{n}\hatvg_{k,0}\nn\\
&\asymp \frac{1}{M^{2}}\tr\bA_{n}^{2}\bD_{k}\bC^{''}\nn\\
&=\frac{ \delta_{k}^{''}}{M}
\end{align}
\begin{align}
\!\!{Q}_{ik}\!\asymp\! \frac{1}{M^{2}}\tr\bR_{k}\bC^{''}\!\!+\!\frac{\left|{\delta_{k}}\right|^{2}\delta_{k}^{''}}{M\left( 1\!+\!\delta_{k} \right)^{2}}\!-\!2\mathrm{Re}\left\{ \! \frac{\delta_{k}\delta_{k}^{''} }{M\left( 1\!+\!\delta_{k} \right)}\!\right\}\!,
 \label{eq:theorem2.I.mu}
\end{align}
where $\bK=\bA_{n}^{2}\bD_{i}$ with  $\delta_{k}^{''}=\frac{1}{M}\tr\bA_{n}^{2}\bD_{k}\bC^{''}$.
This concludes the proof for the derivation of   $\bar{\gamma}_{k,n}$.

\end{appendices}
\section*{Acknowledgement}
The author would like to express his gratitude to Dr. Rajet Krishnan for his help and support in making this work possible.

\bibliographystyle{IEEEtran}

\bibliography{mybib} 

\begin{thebibliography}{10}
\providecommand{\url}[1]{#1}
\csname url@samestyle\endcsname
\providecommand{\newblock}{\relax}
\providecommand{\bibinfo}[2]{#2}
\providecommand{\BIBentrySTDinterwordspacing}{\spaceskip=0pt\relax}
\providecommand{\BIBentryALTinterwordstretchfactor}{4}
\providecommand{\BIBentryALTinterwordspacing}{\spaceskip=\fontdimen2\font plus
\BIBentryALTinterwordstretchfactor\fontdimen3\font minus
  \fontdimen4\font\relax}
\providecommand{\BIBforeignlanguage}[2]{{%
\expandafter\ifx\csname l@#1\endcsname\relax
\typeout{** WARNING: IEEEtran.bst: No hyphenation pattern has been}%
\typeout{** loaded for the language `#1'. Using the pattern for}%
\typeout{** the default language instead.}%
\else
\language=\csname l@#1\endcsname
\fi
#2}}
\providecommand{\BIBdecl}{\relax}
\BIBdecl

\bibitem{Papazafeiropoulos2015b}
\BIBentryALTinterwordspacing
A.~K. Papazafeiropoulos, ``Downlink performance of massive {MIMO} under general
  channel aging conditions,'' in \emph{Proc. IEEE Global Communications
  Conference (GLOBECOM 2015)}, San Diego, USA, December 2015. [Online].
  Available: \url{http://arxiv.org/abs/1509.04303}
\BIBentrySTDinterwordspacing

\bibitem{METIS}
\BIBentryALTinterwordspacing
``{FP7} integrating project {METIS} ({lCT} 317669).'' [Online]. Available:
  \url{https://www.metis2020.com}
\BIBentrySTDinterwordspacing

\bibitem{Marzetta2010}
T.~Marzetta, ``Noncooperative cellular wireless with unlimited numbers of base
  station antennas,'' \emph{IEEE Trans. Wireless Commun.}, vol.~9, no.~11, pp.
  3590--3600, November 2010.

\bibitem{Rusek2013}
F.~Rusek, D.~Persson, B.~K. Lau, E.~Larsson, T.~Marzetta, O.~Edfors, and
  F.~Tufvesson, ``Scaling up {MIMO}: Opportunities and challenges with very
  large arrays,'' \emph{IEEE Signal Processing Mag.}, vol.~30, no.~1, pp.
  40--60, Jan 2013.

\bibitem{5g}
\BIBentryALTinterwordspacing
``The 5{G} infrastructure public private partnership.'' [Online]. Available:
  \url{http://5g-ppp.eu/}
\BIBentrySTDinterwordspacing

\bibitem{Larsson2014}
E.~Larsson, O.~Edfors, F.~Tufvesson, and T.~Marzetta, ``Massive {MIMO} for next
  generation wireless systems,'' \emph{IEEE Commun. Mag.}, vol.~52, no.~2, pp.
  186--195, February 2014.

\bibitem{Osseiran2014}
A.~Osseiran, F.~Boccardi, V.~Braun, K.~Kusume, P.~Marsch, M.~Maternia,
  O.~Queseth, M.~Schellmann, H.~Schotten, H.~Taoka, H.~Tullberg, M.~Uusitalo,
  B.~Timus, and M.~Fallgren, ``Scenarios for 5{G} mobile and wireless
  communications: The vision of the {METIS} project,'' \emph{IEEE Commun.
  Mag.}, vol.~52, no.~5, pp. 26--35, May 2014.

\bibitem{Couillet2011}
R.~Couillet and M.~Debbah, \emph{Random matrix methods for wireless
  communications}.\hskip 1em plus 0.5em minus 0.4em\relax Cambridge University
  Press, 2011.

\bibitem{Gesbert2007}
D.~Gesbert, M.~Kountouris, R.~Heath, C.-B. Chae, and T.~Salzer, ``Shifting the
  {MIMO} paradigm,'' \emph{IEEE Signal Process. Mag.}, vol.~24, no.~5, pp.
  36--46, Sept 2007.

\bibitem{Bjoernson2015}
\BIBentryALTinterwordspacing
E.~Bj{\"o}rnson, E.~G. Larsson, and T.~L. Marzetta, ``Massive {MIMO}: 10 myths
  and one grand question,'' 2015. [Online]. Available:
  \url{http://arxiv.org/abs/1503.06854}
\BIBentrySTDinterwordspacing

\bibitem{Truong2013}
K.~Truong and R.~Heath, ``Effects of channel aging in massive {MIMO} systems,''
  \emph{IEEE/KICS Journal of Communications and Networks, Special Issue on
  massive {MIMO}}, vol.~15, no.~4, pp. 338--351, Aug 2013.

\bibitem{Papazafeiropoulos2014}
A.~Papazafeiropoulos and T.~Ratnarajah, ``Uplink performance of massive {MIMO}
  subject to delayed {CSIT} and anticipated channel prediction,'' in \emph{IEEE
  International Conference on Acoustics, Speech and Signal Processing (ICASSP),
  2014}, May 2014, pp. 3162--3165.

\bibitem{Papazafeiropoulos2014WCNC}
------, ``Linear precoding for downlink massive {MIMO} with delayed {CSIT} and
  channel prediction,'' in \emph{IEEE Wireless Communications and Networking
  Conference (WCNC), 2014}, April 2014, pp. 809--914.

\bibitem{Papazafeiropoulos2015a}
A.~K. Papazafeiropoulos and T.~Ratnarajah, ``Deterministic equivalent
  performance analysis of time-varying massive {MIMO} systems,'' \emph{IEEE
  Trans. on Wireless Commun.}, vol.~14, no.~10, pp. 5795--5809, 2015.

\bibitem{Papazafeiropoulos2015}
\BIBentryALTinterwordspacing
A.~K. Papazafeiropoulos, ``Impact of user mobility on optimal linear receivers
  in cellular networks,'' in \emph{2015 {IEEE} International Conference on
  Communications, {ICC} 2015, London, United Kingdom, June 8-12, 2015}, 2015,
  pp. 2239--2244. [Online]. Available:
  \url{http://dx.doi.org/10.1109/ICC.2015.7248658}
\BIBentrySTDinterwordspacing

\bibitem{PapazafeiropoulosPIMRC}
A.~K. Papazafeiropoulos, H.~Q. Ngo, M.~Matthaiou, and T.~Ratnarajah, ``Uplink
  performance of conventional and massive {MIMO} cellular systems with delayed
  {CSIT},'' in \emph{IEEE International Symposium on Personal, Indoor and
  Mobile Radio Communications (PIMRC), 2014}, Washington, D.C., September 2014,
  pp. 574--579.

\bibitem{Kong2015}
C.~Kong, C.~Zhong, A.~K. Papazafeiropoulos, M.~Matthaiou, and Z.~Zhang,
  ``Sum-rate and power scaling of massive {MIMO} systems with channel aging,''
  \emph{IEEE Trans. on Commun.}, vol.~63, no.~12, pp. 4879--4893, 2015.

\bibitem{Kong2015a}
------, ``Effect of channel aging on the sum-rate of uplink massive {MIMO}
  systems,'' in \emph{IEEE International Symposium on Information Theory (ISIT)
  2015}.\hskip 1em plus 0.5em minus 0.4em\relax IEEE, 2015, pp. 1222--1226.

\bibitem{Schenk2008}
T.~Schenk, \emph{RF imperfections in high-rate wireless systems: impact and
  digital compensation}.\hskip 1em plus 0.5em minus 0.4em\relax Springer
  Science \& Business Media, 2008.

\bibitem{Studer2010}
C.~Studer, M.~Wenk, and A.~Burg, ``{MIMO} transmission with residual
  transmit-{RF} impairments,'' in \emph{ITG/IEEE Work. Smart Ant. (WSA)}.\hskip
  1em plus 0.5em minus 0.4em\relax IEEE, 2010, pp. 189--196.

\bibitem{Qi2012}
J.~Qi and S.~A{\"\i}ssa, ``On the power amplifier nonlinearity in {MIMO}
  transmit beamforming systems,'' \emph{IEEE Trans. Commun.}, vol.~60, no.~3,
  pp. 876--887, 2012.

\bibitem{Mehrpouyan2012}
H.~Mehrpouyan, A.~Nasir, S.~Blostein, T.~Eriksson, G.~Karagiannidis, and
  T.~Svensson, ``Joint estimation of channel and oscillator phase noise in
  {MIMO} systems,'' \emph{IEEE Trans. Signal Process.}, vol.~60, no.~9, pp.
  4790--4807, Sept 2012.

\bibitem{Goransson2008}
B.~Goransson, S.~Grant, E.~Larsson, and Z.~Feng, ``Effect of transmitter and
  receiver impairments on the performance of {MIMO} in {HSDPA},'' in \emph{IEEE
  9th Int. Workshop Signal Process. Adv. Wireless Commun. (SPAWC)}, 2008, pp.
  496--500.

\bibitem{Bjornson2012Optimal}
E.~Bj{\"o}rnson, P.~Zetterberg, and M.~Bengtsson, ``Optimal coordinated
  beamforming in the multicell downlink with transceiver impairments,'' in
  \emph{IEEE Global Commun.Conf. (GLOBECOM), 2012}, Dec 2012, pp. 4775--4780.

\bibitem{Qi2010}
J.~Qi and S.~A{\"\i}ssa, ``Analysis and compensation of {I/Q} imbalance in
  {MIMO} transmit-receive diversity systems,'' \emph{IEEE Trans. Commun.},
  vol.~58, no.~5, pp. 1546--1556, 2010.

\bibitem{Bjoernson2013}
E.~Bj{\"o}rnson, P.~Zetterberg, M.~Bengtsson, and B.~Ottersten, ``Capacity
  limits and multiplexing gains of {MIMO} channels with transceiver
  impairments,'' \emph{IEEE Commun. Lett.}, vol.~17, no.~1, pp. 91--94, 2013.

\bibitem{Zhang2015}
W.~Zhang, H.~Ren, C.~Pan, M.~Chen, R.~C. de~Lamare, B.~Du, and J.~Dai,
  ``Large-scale antenna systems with {UL/DL} hardware mismatch: {A}chievable
  rates analysis and calibration,'' \emph{IEEE Trans. on Commun.}, vol.~63,
  no.~4, pp. 1216--1229, 2015.

\bibitem{Pitarokoilis2015}
A.~Pitarokoilis, S.~Mohammed, and E.~Larsson, ``Uplink performance of
  time-reversal {MRC} in massive {MIMO} systems subject to phase noise,''
  \emph{IEEE Trans. Wireless Commun.}, vol.~14, no.~2, pp. 711--723, Feb 2015.

\bibitem{Bjornson2015}
E.~Bjornson, M.~Matthaiou, and M.~Debbah, ``Massive {MIMO} with non-ideal
  arbitrary arrays: Hardware scaling laws and circuit-aware design,''
  \emph{IEEE Trans. on Wireless Commun.,}, vol.~14, no.~8, pp. 4353--4368,
  2015.

\bibitem{Krishnan2015}
\BIBentryALTinterwordspacing
R.~Krishnan, M.~Khanzadi, N.~Krishnan, Y.~Wu, A.~Amat, T.~Eriksson, and
  R.~Schober, ``Linear massive {MIMO} precoders in the presence of phase
  noise--{A} large-scale analysis,'' \emph{accepted in IEEE Trans. on Veh.
  Tech.}, 2015. [Online]. Available: \url{http://arxiv.org/abs/1501.05461}
\BIBentrySTDinterwordspacing

\bibitem{Bjornson2014}
E.~Bj{\"o}rnson, J.~Hoydis, M.~Kountouris, and M.~Debbah, ``Massive {MIMO}
  systems with non-ideal hardware: Energy efficiency, estimation, and capacity
  limits,'' \emph{IEEE Trans. Inform. Theory}, vol.~60, no.~11, pp. 7112--7139,
  Nov 2014.

\bibitem{Demir2000}
A.~Demir, A.~Mehrotra, and J.~Roychowdhury, ``Phase noise in oscillators: A
  unifying theory and numerical methods for characterization,'' \emph{IEEE
  Trans. Circuits Syst. {I}}, vol.~47, no.~5, pp. 655--674, May 2000.

\bibitem{Bittner2007}
S.~Bittner, E.~Zimmermann, and G.~Fettweis, ``Iterative phase noise mitigation
  in {MIMO-OFDM} systems with pilot aided channel estimation,'' in \emph{IEEE
  66th Vehicular Technology Conference, 2007. VTC-2007 Fall. 2007}, Sept 2007,
  pp. 1087--1091.

\bibitem{Wu2004}
S.~Wu and Y.~Bar-Ness, ``{OFDM} systems in the presence of phase noise:
  Consequences and solutions,'' \emph{IEEE Trans. Commun.}, vol.~52, no.~11,
  pp. 1988--1996, Nov 2004.

\bibitem{Petrovic2007}
D.~Petrovic, W.~Rave, and G.~Fettweis, ``Effects of phase noise on {OFDM}
  systems with and without {PLL}: Characterization and compensation,''
  \emph{IEEE Trans. Commun.}, vol.~55, no.~8, pp. 1607--1616, Aug 2007.

\bibitem{Krishnan2014}
\BIBentryALTinterwordspacing
R.~Krishnan, M.~R. Khanzadi, N.~Krishnan, A.~Amat, T.~Eriksson, N.~Mazzali, and
  G.~Colavolpe, ``On the impact of oscillator phase noise on the uplink
  performance in a massive {MIMO-OFDM} system,'' \emph{Under Review in IEEE
  Signal Processing Letters}, 2014. [Online]. Available:
  \url{http://arxiv.org/abs/1405.0669}
\BIBentrySTDinterwordspacing

\bibitem{Pitarokoilis2013}
A.~Pitarokoilis, S.~Mohammed, and E.~Larsson, ``Achievable rates of {ZF}
  receivers in massive {MIMO} with phase noise impairments,'' in \emph{Asilomar
  Conference on Signals, Systems and Computers, 2013}, Nov 2013, pp.
  1004--1008.

\bibitem{Ngo2013}
H.~Q. Ngo, E.~Larsson, and T.~Marzetta, ``Energy and spectral efficiency of
  very large multiuser {MIMO} systems,'' \emph{IEEE Trans. Commun.}, vol.~61,
  no.~4, pp. 1436--1449, April 2013.

\bibitem{Hoydis2013}
J.~Hoydis, S.~Ten~Brink, and M.~Debbah, ``Massive {MIMO} in the {UL/DL} of
  cellular networks: How many antennas do we need?'' \emph{IEEE J. Select.
  Areas Commun.}, vol.~31, no.~2, pp. 160--171, February 2013.

\bibitem{Truong2014}
K.~T. Truong, A.~Lozano, and R.~W. Heath, ``Optimal training in continuous
  flat-fading massive {MIMO} systems,'' in \emph{Proceedings of 20th European
  Wireless Conference European Wireless 2014}, May 2014, pp. 1--6.

\bibitem{Ying2014}
D.~Ying, F.~Vook, T.~Thomas, and D.~Love, ``Sub-sector-based codebook feedback
  for massive {MIMO} with {2D} antenna arrays,'' in \emph{IEEE Global
  Communications Conference (GLOBECOM), 2014}, Dec 2014, pp. 3702--3707.

\bibitem{Vu2007}
M.~Vu and A.~Paulraj, ``On the capacity of {MIMO} wireless channels with
  dynamic {CSIT},'' \emph{IEEE J. Select. Areas Commun.}, vol.~25, no.~7, pp.
  1269--1283, September 2007.

\bibitem{WCJr1974}
W.~C. Jakes, \emph{Microwave mobile communications}.\hskip 1em plus 0.5em minus
  0.4em\relax New York: Wiley, 1974.

\bibitem{Medard2000}
M.~Medard, ``The effect upon channel capacity in wireless communications of
  perfect and imperfect knowledge of the channel,'' \emph{IEEE T. Inform.
  Theory}, vol.~46, no.~3, pp. 933--946, May 2000.

\bibitem{Hassibi2003}
B.~Hassibi and B.~Hochwald, ``How much training is needed in multiple-antenna
  wireless links?'' \emph{IEEE Trans. Inform. Theory}, vol.~49, no.~4, pp.
  951--963, April 2003.

\bibitem{Billingsley2008}
P.~Billingsley, \emph{Probability and measure}, 3rd~ed.\hskip 1em plus 0.5em
  minus 0.4em\relax John Wiley \& Sons, Inc., 2008.

\bibitem{Vaart2000}
V.~d. Vaart and W.~Aad, \emph{Asymptotic statistics (Cambridge series in
  statistical and probabilistic mathematics)}.\hskip 1em plus 0.5em minus
  0.4em\relax New York: Cambridge University Press, 2000.

\bibitem{Colavolpe2005}
G.~Colavolpe, A.~Barbieri, and G.~Caire, ``Algorithms for iterative decoding in
  the presence of strong phase noise,'' \emph{IEEE J. Select. Areas Commun.},
  vol.~23, no.~9, pp. 1748--1757, Sept 2005.

\bibitem{Bai1}
J.~W. Silverstein and Z.~Bai, ``On the empirical distribution of eigenvalues of
  a class of large dimensional random matrices,'' \emph{Journal of Multivariate
  analysis}, vol.~54, no.~2, pp. 175--192, 1995.

\bibitem{Bai2}
Z.~Bai, J.~W. Silverstein \emph{et~al.}, ``On the signal-to-interference ratio
  of {CDMA} systems in wireless communications,'' \emph{The Annals of Applied
  Probability}, vol.~17, no.~1, pp. 81--101, 2007.

\bibitem{Tao2012}
T.~Tao, \emph{Topics in random matrix theory}.\hskip 1em plus 0.5em minus
  0.4em\relax American Mathematical Soc., 2012, vol. 132.

\bibitem{Wagner2012}
S.~Wagner, R.~Couillet, M.~Debbah, and D.~Slock, ``Large system analysis of
  linear precoding in correlated {MISO} broadcast channels under limited
  feedback,'' \emph{IEEE Trans. Inform. Theory}, vol.~58, no.~7, pp.
  4509--4537, July 2012.

\bibitem{Kay}
S.~M. Kay, \emph{Fundamentals of statistical signal processing: Estimation
  theory}.\hskip 1em plus 0.5em minus 0.4em\relax Upper Saddle River: Prentice
  Hall PTR, 1993.

\end{thebibliography}
\begin{IEEEbiography}[{\includegraphics[width=1in,height=1.25in,clip,keepaspectratio]{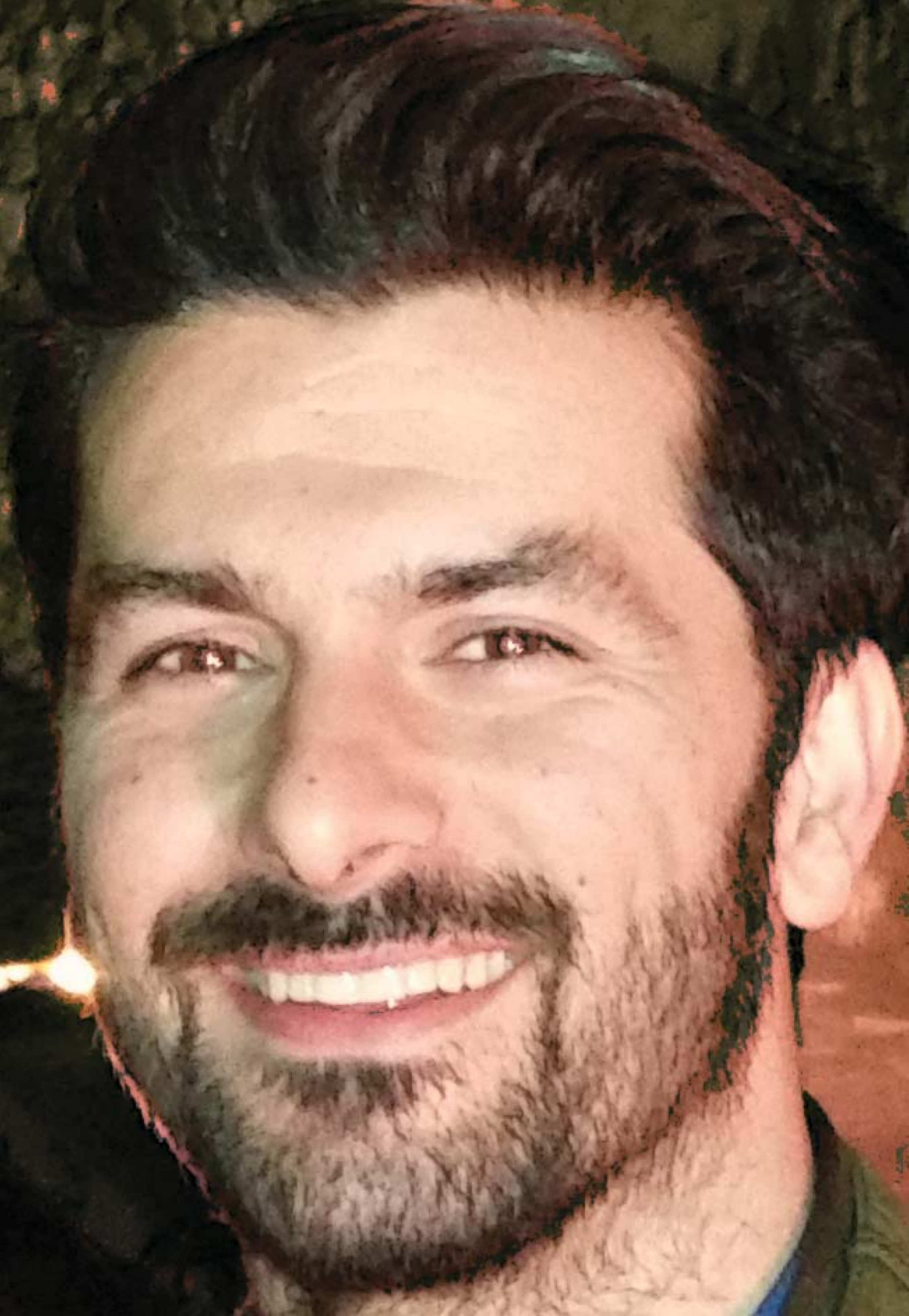}}]{Anastasios Papazafeiropoulos}is currently a Research Fellow in IDCOM at the University of Edinburgh, U.K. He obtained the B.Sc in physics and the M.Sc. in electronics and computers science both with distinction from the University of Patras, Greece in 2003 and 2005, respectively. He then received the Ph.D. degree from the same university in 2010. From November 2011 through December 2012 he was with the Institute for Digital Communications (IDCOM) at the University of Edinburgh, U.K. working as a postdoctoral Research Fellow, while during 2012-2014 he was a Marie Curie Fellow at Imperial College London, U.K.   His research interests span massive MIMO, 5G wireless networks, full-duplex radio, mmWave communications, random matrices theory, signal processing for wireless communications, hardware-constrained communications,
and performance analysis of fading channels. 
\end{IEEEbiography}

\end{document}